\documentclass[twocolumn, nofootinbib, aps, prd, 10pt]{revtex4-1}
\usepackage{graphicx}
\usepackage{float}
\usepackage{url}
\usepackage{color}
\usepackage{rotating}
\usepackage{amssymb,amsmath}
\usepackage{mathtools} 
\usepackage{booktabs}
\pdfoutput=1

\newcommand{\alm}{a_{\ell m}}
\newcommand{\glm}{g_{\ell m}}

\newcommand{\ISW}{s} 
\newcommand{\gal}{g}
\newcommand{\cmb}{T}
\newcommand{\Clss}{\Cl^{\ISW\ISW}}
\newcommand{\Clgs}{\Cl^{\ISW\gal}}
\newcommand{\Clgg}{\Cl^{\gal\gal}}
\newcommand{\Cltt}{\Cl^{\cmb\cmb}}
\newcommand{\Clpp}{\Cl^{pp}}
\newcommand{\Dl}{D_\ell}

\newcommand{\tClrec}{\tilde \Cl^{\rm rec-rec}}

\newcommand{\Cl}{C_\ell}

\newcommand{\Clcal}{C_{\ell}^{\rm cal}}
\newcommand{\Clfid}{C_{\ell}^{\rm th}}
\newcommand{\Cltrue}{C_{\ell}^{\rm sky}}
\renewcommand{\vec}[1]{\boldmath{#1}}

\newcommand{\Var}{\rm Var}

\newcommand{\nhat}{{\bf \hat n}}
\newcommand{\Rl}{R_\ell}
\newcommand{\Rfid}{{R_\ell(\Cl^{\rm th})}}
\newcommand{\Rtrue}{{R_\ell(\Cl^{\rm sky})}}

\newcommand{\varc}{{\sigma^2_{\rm cal}}}

\newcommand\myeqthree{\stackrel{\mathclap{\normalfont\mbox{\scriptsize single map}}}{\hspace{0.45cm}\longrightarrow\hspace{0.45cm}}}

\newcommand{\Nobs}{N_{\rm obs}}

\everymath{\displaystyle}
  
\newcommand{\be}{\begin{equation}}
\newcommand{\ee}{\end{equation}}

\begin{document}

\title{Integrated Sachs-Wolfe map reconstruction in the presence of systematic errors}

\author{Noah Weaverdyck,$^*$ Jessica Muir and Dragan Huterer}
\affiliation{Department of Physics, University of Michigan, 
450 Church St, Ann Arbor, MI 48109-1040}
\email[Corresponding author: ]{nweaverd@umich.edu}
\date{\today}

\begin{abstract}
The decay of gravitational potentials in the presence of dark energy leads to
an additional, late-time contribution to anisotropies in the cosmic microwave
background (CMB) at large angular scales. The imprint of this so-called
integrated Sachs-Wolfe (ISW) effect to the CMB angular power spectrum has been
detected and studied in detail, but reconstructing its spatial contributions
to the CMB \textit{map}, which would offer the tantalizing possibility of
separating the early- from the late-time contributions to CMB temperature
fluctuations, is more challenging. Here we study the technique for
reconstructing the ISW map based on information from galaxy surveys and focus
in particular on how its accuracy is impacted by the presence of photometric
calibration errors in input galaxy maps, which were previously found to be a
dominant contaminant for ISW signal estimation. We find that both including
tomographic information from a single survey and using data from multiple,
complementary galaxy surveys improve the reconstruction by 
mitigating the impact of
spurious power contributions from calibration errors. A high-fidelity
reconstruction further requires one to account for the contribution of
calibration errors to the observed galaxy power spectrum in the model used to
construct the ISW estimator.  We find that if the photometric calibration
errors in galaxy surveys can be independently controlled at the level required
to obtain unbiased dark energy constraints, then it is possible to reconstruct
ISW maps with excellent accuracy using a combination of maps from two galaxy
surveys with properties similar to Euclid and SPHEREx.
\end{abstract}
\maketitle
\section{Introduction}\label{sec:intro}

Cosmic microwave background (CMB) photons undergo a frequency shift as they
travel to us from the last scattering surface. On top of the redshift due to the
expansion of the Universe,  an additional contribution to the
temperature anisotropy is introduced  whenever the universe is not matter
dominated---for example, right after recombination when radiation
contributes non-negligibly, or at late times when dark energy becomes
important.  This so-called integrated Sachs-Wolfe (ISW) effect  is given by
\citep{Sachs:1967er,Hu:1993xh} 
\begin{equation}\label{ISWorigexpr}
  \left.\frac{\Delta T}{\bar{T}}\right|_{ISW}(\nhat) =
  \frac{2}{c^2}\int_{t_*}^{t_0}dt\,\frac{\partial \Phi(\mathbf{r},t)}{\partial t},
\end{equation}
where $t_0$ is the present time, $t_{\star}$ is that of recombination, $c$ is
the speed of light, $\mathbf{r}$ is the position in comoving coordinates, and
$\Phi$ is the gravitational potential.  The late-time ISW signal 
(referred to hereafter simply as `ISW')
 has been
statistically detected via measurements of the cross-correlation of CMB
temperature maps with galaxy maps \citep{Boughn:2003yz,Fosalba:2003ge,
  Nolta:2003uy, Corasaniti:2005pq, Padmanabhan:2004fy, Vielva:2004zg,
  McEwen:2006my,Giannantonio:2006du, Cabre:2007rv, Rassat:2006kq,
  Giannantonio:2008zi,Ho:2008bz,Xia:2009dr,Giannantonio:2012aa,Ade:2013dsi,Ade:2015dva}
and, more recently, with maps of CMB lensing convergence 
\cite{Ade:2013dsi,Ade:2015dva}. These detections serve as an important
consistency test of the standard model of cosmology, and can help constrain
the properties of dark energy.

The ISW can provide additional information beyond its power spectrum
if its {\it map} can be reconstructed with sufficient signal-to-noise.  
 Since the total large-angle CMB temperature anisotropy is the sum of early- (hereafter `primordial')
 and late-time contributions,
\begin{equation}
  \left.\frac{\Delta T}{\bar{T}}\right|(\nhat) = 
  \left.\frac{\Delta T}{\bar{T}}\right|_\mathrm{prim}(\nhat) + 
  \left.\frac{\Delta T}{\bar{T}}\right|_\mathrm{ISW}(\nhat),
\end{equation}
reconstructing the ISW map would allow us to isolate the primordial-only
anisotropy. This separation of the CMB into
early- and late-time contributions can also be useful for a variety
of cosmological tests. For example, one could study the temporal origin of the
large-angle CMB anomalies reported in, e.g., Ref~\cite{Schwarz:2015cma}. One
could also subtract the realization-specific contaminating ISW contribution to
estimation of primordial non-Gaussianity \cite{Kim:2013nea}, something that is
currently done using theoretical templates for the ISW-lensing bispectrum
\cite{Ade:2015ava}. Motivated by these considerations, reconstruction of the ISW map has been
the focus of a number of recent
analyses~\cite{Barreiro:2008sn,Granett:2008dz,Francis:2009pt,Barreiro:2012yh,Ade:2013dsi,Rassat:2013caa,Manzotti:2014kta,Granett:2015dna,Ade:2015dva,Muir:2016veb,Bonavera:2016hbm}.

In this paper we study how ISW map reconstruction is affected by a class of
observational and astrophysical systematic errors 
which we will refer to broadly as photometric calibration
errors or, for conciseness, calibration errors. These systematics afflict all
galaxy surveys at large angular scales, contributing to the significant excess of power at large scales found in many recent surveys, including the Sloan Digital Sky Survey (SDSS)
\cite{Giannantonio:2012aa,Pullen:2012rd, Ho:2012vy, Ho:2013lda,
  Agarwal:2013qta,Giannantonio:2013uqa,
  Agarwal:2013ajb,Hernandez-Monteagudo:2013vwa}, MegaZ \cite{Thomas:2010tn},
WISE-AGN and WISE-GAL \cite{Ade:2013dsi}, and NVSS
\cite{Blake:2001bg,HernandezMonteagudo:2009fb,Ho:2008bz,Giannantonio:2012aa}.
Calibration errors are thus already established as one of the most
significant  systematics impacting large-angle measurements of
galaxy surveys, a fact that has broad implications, such as for measuring
scale-dependent bias as a signal of primordial non-Gaussianity. As the
statistical power of galaxy surveys continues to grow, the control and understanding
of systematics like calibration errors is becoming even more important.

There is a variety of ways in which modern photometric surveys assess and
mitigate contamination from systematics, many of which rely on
cross-correlating galaxy maps with known systematics templates. This can be
used to identify contaminated regions, which are then masked or excluded from
the analysis (as in Ref.~\cite{Scranton:2001xa}). 
Alternatively, one can use these templates to subtract or marginalize
over systematics-induced spatial variations in the calculation of, for
example, the two-point clustering signal \cite{Ross:2011cz, Leistedt:2013gfa,
  Leistedt:2014wia, Elsner:2015aga, Hernandez-Monteagudo:2013vwa,
  Awan:2016zuk, Elsner:2016bvs}.  Such an approach was taken in
Ref.~\cite{Hernandez-Monteagudo:2013vwa} to study the overall detection
significance of the ISW effect in SDSS data. The authors found results similar to
Ref.~\citep{Giannantonio:2012aa}, the authors of which instead accounted for excess power by adding
a low-redshift spike in the source distributions.  Most of these correlation
corrections are perturbative, however, and therefore require
fairly clean maps in which systematic effects are minor to begin
with. Additionally, while corrections to the two-point statistics are important
 for the inference of cosmological parameters, they do not remove the systematics from the maps themselves. 
\citet{Suchyta:2015qbs} propose an alternative approach, wherein 
 measurement biases are characterized by injecting fake objects into Dark
Energy Survey images. This neatly avoids the reliance on having small levels of contamination
in the input maps, but it still cannot account for certain systematics, such as dust or 
flux calibration. 
Whatever the approach taken, some level of residual calibration error will remain.

Some of us previously showed  that at levels of calibration control 
consistent with current and near-future
surveys, residual
calibration errors are by far the dominant systematic for ISW signal
reconstruction~\cite{Muir:2016veb}.  This motivates us to study their impact in more detail.
Namely, we would like to study whether the presence of residual calibration
errors can be mitigated by combining information from multiple input maps 
 or through better modeling of the contributions of systematics to observed galaxy power. We also wish to
investigate to what extent residual calibration errors similarly impact the
signal-to-noise ratio of galaxy-CMB cross-correlation and, in turn, the significance of ISW detection.  
With this aim, we use ensembles of simulated maps to characterize the performance of ISW
reconstruction based on surveys like
Euclid and SPHEREx, two proposed wide-angle surveys of which the properties are
expected to be good for ISW detection and reconstruction.  We also consider
the benefits of including Planck-like simulations of CMB intensity in the
reconstruction effort.

We begin in Sec.~\ref{sec:method} by describing our model for
 calibration errors, how we reconstruct the ISW map and evaluate its quality, and which
  input data sets we use. In Sec.~\ref{sec:results}, we compare the performance of ISW reconstruction
   when using one versus multiple surveys and investigate the impact  different assumptions have
   on the results. In Sec.~\ref{sec:SNcal}, we relate map reconstruction to the total signal-to-noise ratio of ISW detection, and we conclude in Sec.~\ref{sec:conclusion}.

\begin{figure}
\includegraphics[width=\linewidth]{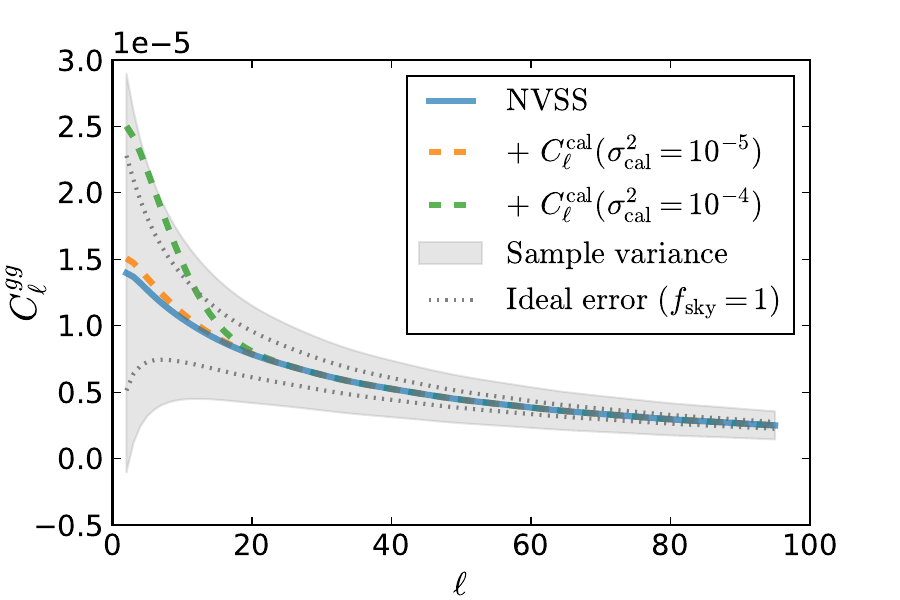}
  \caption{Effects of calibration errors on galaxy power. The solid curve shows the theoretical
    angular power spectrum for the NVSS survey
    \cite{Ade:2013dsi,Ade:2015dva}. The colored dashed curves show the theoretical
    spectrum with two representative levels of calibration error. The shaded region is the $1\sigma$
    uncertainty from the survey's sample variance, and the dotted curves indicate the ideal,
    all-sky cosmic variance.}
      \label{fig:plancknvssplot}
\end{figure}

\begin{figure}
\includegraphics[width=\linewidth]{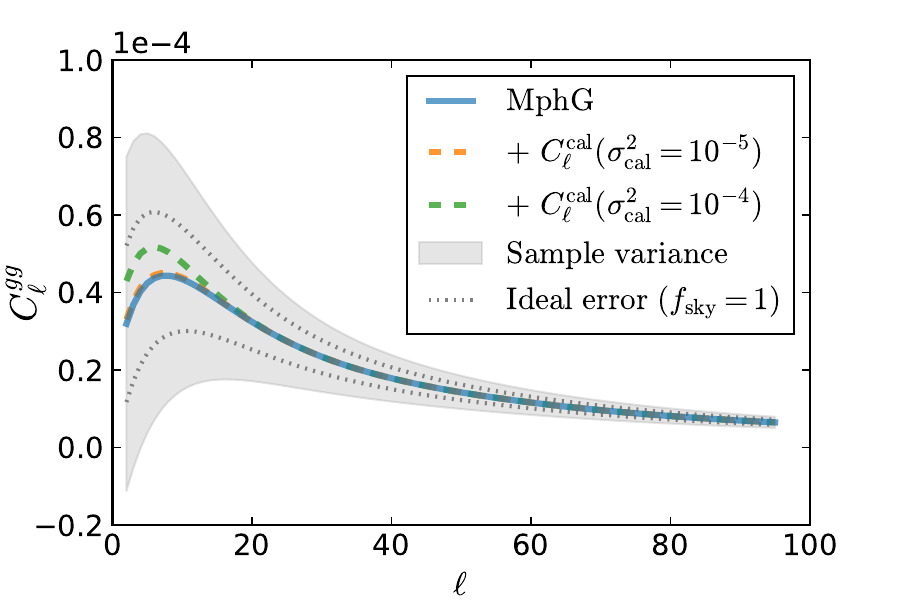}
  \caption{Same as Fig.~\ref{fig:plancknvssplot}, but for the SDSS MphG
    catalog, following Refs.~\cite{Ade:2013dsi,Ade:2015dva}. In this case, the
    sample variance is driven by sky coverage ($f_{sky}$=0.22) as opposed to
   number density as for NVSS.}
  \label{fig:planckmphgplot}
\end{figure}

\section{Methodology}\label{sec:method}
\subsection{Modeling calibration errors}\label{sec:calmodel}

Photometric calibration is a challenge faced by all photometric galaxy surveys. It refers to the adjustments required to establish
a consistent spatial and temporal measurement of flux of the target objects in
different observation bands. A number of different systematics must be
accounted for in calibration, including but not limited to detector
sensitivity variation on the focal plane, variation in observing conditions, the
presence of foreground stars (as galaxies near them are less likely to be
detected), and extinction by dust. Calibration errors are introduced if these systematics are incompletely or inaccurately accounted for.

Our focus is on how calibration errors affect galaxy number counts. To illustrate this, imagine that a perfectly uniform screen
(of e.g.\ dust) blocks some light from all galaxies. This pushes the faintest
galaxies below the survey's flux limit, and leads to observation of fewer
galaxies in all directions. A pure monopole change such as this  increases shot noise
but does not affect the angular clustering signal of galaxies. In contrast, in
a more realistic scenario where the opacity of this ``screen'' depends on
direction, it affects the observed galaxy clustering signal by adding spurious
power and by coupling different multipoles of the measured power
spectrum~\cite{Huterer:2012zs,Shafer:2014wba}. The presence of unaccounted-for
calibration errors can thus introduce biases in cosmological parameter
estimates from Large Scale Structure (LSS) surveys. These are particularly severe for the ISW effect and other 
measurements based
on signals that, like calibration errors, enter primarily at large angular
scales.

In keeping with the picture of  calibration errors as a
direction-dependent screen, we model them as a modulation of the true galaxy
number counts $N(\hat{\bf n})$, where $\nhat$ is the direction on
the sky. The observed, modulated counts are \cite{Huterer:2012zs}
\begin{equation}
\Nobs({\hat{\bf n}}) = [1+c({\hat{\bf n}})]\, N({\hat{\bf n}}), \label{eq:nobs}
\end{equation}
where the field $c({\hat{\bf n}})$ describes the screening effect of
calibration errors.  While we will generally refer to this kind of modulation as the result of  ``calibration errors," as
Eq.~(\ref{eq:nobs}) makes clear, this formalism can describe any residual
effect that modulates a survey's selection function. 

 Though the expression in Eq.~(\ref{eq:nobs}) will necessarily couple different
multipoles, at low $\ell$, the impact of calibration errors on the observed
galaxy power spectrum is well approximated by
\be
  \Cl^{\rm obs}\approx \Cl^{\rm gal} + \Clcal, 
  \label{eq:clobs}
\ee
 neglecting multiplicative terms.

Following \citet{Muir:2016veb}, we model the calibration error field $c(\nhat)$ as a Gaussian random field with power spectrum $\Clcal$ and quantify the level of residual calibration errors using its variance,
\begin{equation}
\varc \equiv \Var(c(\hat{\bf n})) = \sum_{\ell=0}^\infty {2\ell+1\over 4\pi}\Clcal.
\label{eq:calib_var}
\end{equation}
While the conversion between $\sigma_{\rm cal}$ and the rms magnitude error
  depends on the faint-end slope of the luminosity function of tracers in the
  survey, they can be related roughly as $(\delta m)^{\rm rms}\simeq \sigma_{\rm cal}$
  \cite{Huterer:2012zs}. Thus a survey with residual calibration
  errors of $\varc=10^{-6}$ has been calibrated to roughly a milimagnitude.
  
Motivated by power spectrum estimates for maps of dust extinction corrections
and magnitude limit variations in existing surveys (see Fig.~5 and~6
in Ref.~\cite{Huterer:2012zs}), we choose the fiducial calibration error power
spectrum to be
\begin{equation}\label{eq:clcalmodel}
  \Cl^{\rm cal} = \alpha^{\rm cal} \exp{\left[-(\ell/w^{\rm cal})^2\right]},
\end{equation}
with $w^{\rm cal}=10$. The normalization constant $\alpha^{\rm cal}$
is varied to achieve the desired $\varc$.
 Figures~\ref{fig:plancknvssplot} and \ref{fig:planckmphgplot} show
  the impact of calibration errors of this form on the
 angular power spectrum of the NVSS and SDSS MphG galaxy surveys, 
 which have been used to reconstruct the ISW map in previous studies \cite{Ade:2013dsi,Manzotti:2014kta,Ade:2015dva}

For our fiducial model, we assume that calibration error maps for different redshift bins and surveys are uncorrelated with one another. We briefly examine the impact of relaxing such an
assumption in Sec.~\ref{sec:caldetails}.

\subsection{ISW estimator}\label{sec:iswest}

Similarly to~\citet{Muir:2016veb}, we work with the optimal estimator derived by~\citet{Manzotti:2014kta}.  It takes as input $n$ maps, which can 
include any tracers that carry information about the ISW signal, namely LSS,
CMB, or lensing convergence maps. Letting $\glm^i$ represent the spherical
components of the $i$th input map, where $i\in\{1,\dots,n\}$, the optimal
estimator for the spherical component of the ISW signal is 
\be\label{eq:iswest_simple}
\hat{a}_{\ell m}^{\rm ISW} = \sum_{i=1}^nR_{\ell}^i\,\glm^i. 
\ee
The operator
\begin{equation}
  R_{\ell}^i\equiv -N_{\ell}[D_{\ell}^{-1}]_{{\rm ISW-}i}
  \label{eq:R_def}
\end{equation}
is a reconstruction filter derived
from the covariance matrix,
\be
\label{eq:Dl}
D_{\ell} = \left(\begin{array}{cccc}
  C_{\ell}^{\rm{ISW,ISW}}&C_{\ell}^{\rm{ISW,}1} &\cdots &C_{\ell}^{\mathrm{ISW,} n}\\
  C_{\ell}^{1\rm{,ISW}}&C_{\ell}^{1,1} &\cdots &C_{\ell}^{1,n}\\
  \vdots&\vdots &\ddots &\vdots\\
  C_{\ell}^{n\rm{,ISW}}&C_{\ell}^{n,1}&\cdots &C_{\ell}^{n,n}\\
\end{array}\right). 
\ee
In this expression, superscript numbers label the input maps and $N_{\ell} \equiv
1/[D_{\ell}^{-1}]_{\rm ISW-ISW}$ estimates the variance of the reconstruction at
multipole  $\ell$.
When a single input map $A$, is used,  this expression reduces to a simple Wiener filter,
\be
\hat{a}_{\ell m}^{\rm{ISW}} \myeqthree \frac{\Cl^{\rm ISW-A}}{\Cl^{\rm A-A}}\glm^{\rm{A}}.
\label{eq:wiener}
\ee

 We demonstrate in Appendix
  \ref{app:other_est} that Eq.~(\ref{eq:iswest_simple}) is equivalent to the estimator
of \citet{Manzotti:2014kta}, where the CMB temperature map is treated separately from
LSS maps, and show that it reduces to the Linear Covariance Based
(LCB) filter first proposed by~\citet{Barreiro:2008sn}.

In constructing this ISW estimator, one must make a choice about how to obtain the necessary angular power and cross-power spectra in the covariance matrix. 
The $\Cl$'s can either be extracted from observations (as in Refs.~\cite{Francis:2009pt, Rassat:2013caa})
 or computed analytically for an assumed cosmology (as in Refs.~\cite{Ade:2013dsi, Manzotti:2014kta, Ade:2015dva, Muir:2016veb, Bonavera:2016hbm}).
 Analytic calculation is straightforward
 but introduces a model dependence which can potentially bias results if, for example, calibration error contributions are not modeled correctly~\cite{Muir:2016veb}.
 Measuring $\Cl$ from observations produces a model-independent estimator and so can help in the case where the theory spectra are inaccurate,
  but at the expense of limited precision due to sample variance, especially at large scales, scales with low power, or for map combinations
   that have little correlation.\footnote{Using the observed spectra also violates an assumption in the maximum likelihood
    derivation of the estimator, in which the covariance is assumed to be known (i.e. independent of the measured signal).}
     Hybrid methods can also be used, as in Ref.~\cite{Barreiro:2012yh}, where \citet{Barreiro:2012yh} account for observed excesses in the autopower of NVSS data by using a smoothed fit to data to get the galaxy map's autopower, but analytically compute its cross-correlation with the ISW signal.

We therefore consider two limiting cases of constructing the estimator in order to investigate 
how calibration errors impact the ISW reconstruction:
 \begin{enumerate}
   \item a `worst' case estimator filter, $\Rfid$, where we use the fiducial theory $\Cl$'s in
     the estimator, in which calibration errors' power contributions are not
     modeled at all, and
   \item a `best' case estimator filter, $\Rtrue$, in which calibration error power contributions are modeled
     perfectly (i.e. the covariance matrix is known). This case may be approximated by, e.g. a smoothed fit of the
     observed LSS power. 
 \end{enumerate}
 The theoretical spectra are related simply through the expression
   \be
   \label{eq:clcalrel}
   \Cltrue = \Clfid + \Clcal.
   \ee
where $\Clcal$ is the power spectrum of the calibration error field described in Sec.~\ref{sec:calmodel}. 
We consider these cases in Secs.~\ref{sec:euconlyresults} and~\ref{sec:eucpluscmb} respectively.
    
\subsection{Quality statistic}\label{sec:rhodescription}

To quantify the accuracy of a given reconstruction, we use the correlation 
coefficient between the temperature maps of the true [$T^{\rm ISW}(\nhat)$] and
 reconstructed [$T^{\rm rec}(\nhat)$] ISW signal,
\be
\rho = \frac{\frac{1}{N_{\rm pix}} \sum_k^{N_{\rm pix}}
  (T_k^{\rm ISW}-\bar T^{\rm ISW})(T_k^{\rm rec}-\bar T^{\rm rec})}
     {\sigma_{\rm ISW}\sigma_{\rm rec}},
\label{eq:rhofrommap}
\ee
where $\bar{T^{X}}$ and $\sigma_{X}^2$ are the mean and variance of map $T^{X}(\nhat)$, respectively.\footnote{We also considered $s$, which measures the rms error between true and reconstructed ISW maps as a complementary quality statistic, but found that for the cases studied here, the information it provided was largely redundant to that given by $\rho$.}
We do not include pixel weights in our calculation of $\rho$, as is done to account for masking effects in Ref.~\cite{Bonavera:2016hbm}. This is because we work with only full-sky maps, as will be discussed in the next section. 

The correlation coefficient can be rewritten in terms of the cross-power
between the true ISW map realization and the input tracers,

\be
\rho =  \frac{\frac{1}{4\pi}\sum_{\ell, i}(2\ell+1)R_{\ell}^i\tilde{\Cl}^{{\rm ISW}-i}} {\sigma_{\rm ISW}\sigma_{\rm rec} },
\label{eq:rhoraw}
\ee
where the tilde denotes pseduo-$\Cl$ measured from a given map realization,
and we have used Eq.~(\ref{eq:iswest_simple}) to write
\begin{align}
\tilde{C}_{\ell}^{\rm ISW-rec}&=\sum_i\frac{1}{2\ell+1}\sum_{m} [\alm^{\rm ISW}]^*R_{\ell}^i g_{\ell m}^i \\
					&= \sum_iR_{\ell}^i\tilde{C}_{\ell}^{{\rm ISW}-i}.
\end{align}

Because the measured correlation coefficient depends on the specific realization, we assess reconstruction accuracy for a given set of input map properties as follows.
We simulate a large number of realizations of correlated maps, then apply the ISW estimator to obtain associated reconstructed ISW maps, and by comparing these with the true ISW maps we obtain a sample distribution for $\rho$. Its mean value $\bar{\rho}$, which in the limit of an infinitely large ensemble will approach an 
 expectation value $\langle \rho\rangle$, provides a statistical measure of how accurately the estimator can reproduce the true ISW signal. 
  Studying how $\bar{\rho}$ changes in response to variations in survey properties and modeling choices therefore 
  allows us to understand which factors are most important for obtaining an accurate ISW reconstruction.

We can avoid the computational cost of generating many simulation ensembles by
noting that we can obtain a good estimate for the expectation value of
$\rho$ if we make the approximation
\begin{align}
   \langle\rho\rangle &=  \left< {\frac{\frac{1}{4\pi} \sum_{\ell, i}(2\ell+1)R_{\ell}^i\tilde{\Cl}^{{\rm ISW}-i}} {\sigma_{\rm ISW}\sigma_{\rm rec} }} \right> \label{eq:rhoexp}\\
       			&\approx \frac{\frac{1}{4\pi} \sum_{\ell, i}(2\ell+1)R_{\ell}^i\Cl^{{\rm ISW}-i}}{\hat \sigma_{\rm ISW}\hat \sigma_{\rm rec}},
 \label{eq:rhoest}
\end{align}
that is, we replace the pseudo-$\Cl$'s with their expectation value across realizations,
$\tilde{\Cl}~\rightarrow~{\Cl}$. We will refer to the quantity in Eq.~(\ref{eq:rhoest}) as $\hat{\rho}$, defining
\begin{align}
  \hat \sigma_{\rm ISW} &= \sqrt{ \frac{1}{4\pi}\sum_{\ell}\,(2\ell+1)\,\Cl^{\rm ISW}} \\
  \hat \sigma_{\rm rec} &=\sqrt{\frac{1}{4\pi} \sum_{\ell, i,
      j}\,(2\ell+1)\,R_{\ell}^iR_{\ell}^j\Cl^{ij}} \,, \label{eq:sigrecest}
\end{align}
to approximate the rms fluctuations in the true and reconstructed ISW maps. Here
the indices $i$ and $j$ label the input tracer maps and the sum over $\ell$ runs
over the multipoles $\ell \in [2, 95]$, a range chosen to conservatively to include all scales where the ISW signal is important.

We have tested the approximation $\hat{\rho}\approx\langle \rho\rangle$ in Eq.~(\ref{eq:rhoest}) extensively and found it works well
when the estimator filter $R_{\ell}$ is built from analytically computed spectra but
can break down if $R_{\ell}$ is composed of $\tilde\Cl$'s extracted from map realizations. 
This behavior is related to the way in which using measured $\Cl$'s
makes $\rho$ depend on $\tilde\Cl$, such
 that $\bar{\rho} = \rho(\langle\tilde\Cl\rangle)$ is no longer a good approximation of $\langle\rho(\tilde\Cl)\rangle$. Appendix~\ref{sec:appapprox} discusses this in more detail.
 
\subsection{Simulated surveys} \label{sec:modsurveys}

By working with simulated maps, we are able to study in detail how calibration
error levels and modeling choices affect ISW signal reconstruction. 

Since we are concerned only with large scales, we model the ISW signal, total CMB temperature anisotropy, and galaxy number density fluctuations as correlated
Gaussian fields. We use \texttt{HEALPY}~\cite{Gorski:2004by}
                 to generate map realizations based on
 input auto- and cross-power spectra which we compute analytically following the standard expressions given e.g.\ in Ref.~\cite{Muir:2016veb}. We use
 the Limber approximation for $\ell\geq20$, having verified that this affects
 $\rho$  at the level of $0.1\%$ or less for the surveys and range
 of $\varc$ considered here.  We compute $\Cl$ for multipoles with $\ell\leq
 95$, as this range contains almost all of the ISW
 signal~\cite{Afshordi:2004kz}. Accordingly, our simulations are sets of
 \texttt{HEALPIX} maps of resolution ${\rm NSIDE}=(\ell_{\rm max}+1)/3=32$.
 We refer the reader to Ref.~\cite{Muir:2016veb} for a more detailed description of 
 the reconstruction pipeline.

Because our goal is to study the impact of calibration errors and not survey
geometry, we assume full-sky coverage in all of our analyses. 
\citet{Bonavera:2016hbm} found that in overlapping regions of partial sky LSS surveys, 
ISW reconstruction quality degrades only slightly compared to the full-sky case.
Therefore, the performance of a given estimator using full-sky maps should be indicative of its 
performance using maps with only partial sky coverage. 
 
Our fiducial cosmological model
is $\Lambda$CDM, with the best-fit cosmological parameter values from Planck 2015, 
$\{\Omega_ch^2,\Omega_bh^2,\Omega_{\nu}h^2,h,n_s\}=\{0.1188,0.0223,0,0.6774,0.9667\}$. Unless otherwise stated, ISW reconstructions are performed on 2000 map realizations for each analysis and include multipole information down to
$\ell_{\rm min}=2$.

Within this framework, four pieces of information are required to model a LSS survey: the distribution of its sources along the line of sight $n(z)$, a prescription for how they are binned in redshift, their linear bias $b(z)$, and their projected number density per steradian $\bar{n}$. Below we describe how our choices for these characteristics are based on the properties of promising future probes of the ISW effect.

\subsubsection{Euclid-like LSS survey}\label{sec:euclidintro}
Our fiducial survey is modeled on Euclid, a future LSS survey with large 
sky coverage and a deep redshift distribution~\cite{Laureijs:2011gra},  which 
is expected to be an excellent probe of the ISW effect~\cite{Afshordi:2004kz,
Douspis:2008xv}. We assume the redshift distribution used by \citet{Martinet:2015wza}, 
\be\label{eq:fiddndz}
\frac{dn}{dz} = \frac{3}{2z_0^3}z^2\,\exp{\left[-(z/z_0)^{1.5}\right]},
\ee
which has a maximum at $z_{\rm peak}\simeq 1.21z_0$.  We choose $z_0=0.7$ and
$\bar{n} = 3.5\times 10^8$, with a photo-$z$ redshift uncertainty of
$\sigma(z)=0.05(1+z)$ which smoothes the edges of redshift bins. For
simplicity, we assume a constant galaxy bias of $b(z)=1$. Our results 
are qualitatively insensitive to this choice as long as the bias is reasonably
 well approximated for the input maps.  
 This is because the bias term cancels in the estimate of the ISW signal, 
 so that fractional differences between true and modeled bias have little impact on $\rho$.

We refer the reader to Ref.~\cite{Muir:2016veb} for further details on 
both fitting for bias and the impact that mismodeling can have on reconstruction.

In Sec.~\ref{sec:euconlyresults} we investigate the improvement in ISW map
reconstruction when the fiducial Euclid-like survey is split into six
redshift bins with edges at $z\in \{0.01, 0.4, 0.8, 1.2, 1.6, 2,  3.5\}$ (see inset of Fig.~\ref{fig:euconlyplot}), as compared to the unbinned case. We
subsequently use the six-binned Euclid survey as our fiducial case.

\subsubsection{SPHEREx-like LSS survey}\label{sec:spherexintro}

We model a second survey on the SPHEREx All-Sky Spectral Survey (SPHEREx), a proposed survey that has
been optimized to study LSS in the low-redshift universe. One of its goals is to place
stringent limits on primordial non-Gaussianity~\cite{Dore:2014cca},
which will require rigorous control of
calibration errors. Given this, SPHEREx will provide excellent input map
candidates for ISW map reconstruction. Its shallower reach makes it 
complementary to the deeper mapping of the LSS provided by Euclid.

 SPHEREx will identify galaxies with varying levels of redshift uncertainty,
  ranging from  $\sigma_z < 0.003(1+z)$ up to $\sigma_z > 0.1 (1+z)$.
  Grouping these into catalogs with different levels of precision provides collections of galaxies useful for different
science goals.  The $\sigma_z < 0.1(1+z)$ catalog with a projected $\sim 300$
million galaxies was identified in Ref.~\cite{Dore:2014cca} as the best subsample
for $f_{\rm NL}^{\rm loc}$ detection. Our investigations confirm it to be
the best for ISW detection as well.
We therefore fit its projected redshift distribution given in Ref.~\cite{Dore:2014cca} to the functional form for $dn/dz$  given in Eq.~(\ref{eq:fiddndz}). 
We select $z_0=0.46$, which results in a peak $dn/dz$ of $z_{\rm peak} \simeq 0.56$.
We have confirmed that our results are not strongly sensitive to changes in this redshift
distribution, in agreement with the findings of Ref.~\cite{Muir:2016veb}.

We use a projected number density of
$\bar{n} = 6.6\times 10^7$ and consider the case where the survey is split
into six redshift bins. We choose their edges by scaling the Euclid-like survey's binned redshift distribution to the SPHEREx median redshift,  resulting in redshift bin edges at $z \in \{0.01, 0.26,  0.53,  0.79,  1.05, 1.31,  2.30\}$.
This still provides sufficient sampling of the
field in each bin to ensure that shot noise is subdominant to the galaxy signal power.

\subsubsection{Planck-like CMB survey}\label{sec:planckintro}

CMB data have frequently been used in conjunction with LSS data for ISW map reconstruction.
Recent examples include Ref.~\cite{Ade:2013dsi}, which used NVSS radio data, the
Planck lensing convergence map, and Planck temperature data. That analysis was subsequently extended to include more LSS tracers in Ref.~\cite{Ade:2015dva}. However, in both of these cases, residual systematics limit the usefulness of lensing data to scales of $\ell \geq 10$ and $\ell \geq 8$, respectively. 
\citet{Bonavera:2016hbm} investigated the usefulness of CMB data for ISW reconstruction using a simulation 
 pipeline similar to ours, finding that both CMB temperature and polarization data
 only modestly improve reconstruction quality but carry a greater benefit when the
 LSS tracers themselves contain less information (due to e.g. noise or other properties of the survey).

It is then natural to ask whether CMB data can help mitigate
the impact of calibration errors in LSS maps. 
We therefore consider CMB temperature as an additional input map. To compute
the total CMB temperature power spectrum, $\Cl^{TT}$, we compute the
primordial-only contributions using a modified version of
\texttt{CAMB} \cite{Lewis:1999bs} and add them to our calculations for $\Cl^{\rm
  ISW}$. 
  As the CMB power spectrum is determined within the limits of 
cosmic variance at low $\ell$ and the ISW signal is already dominated by the primary (that is,
non-ISW) CMB anisotropies, we do not include calibration errors in the generation of CMB temperature maps.
 Though CMB polarization and lensing could provide
additional information, residual systematics remain at large scales for each 
(see Refs.~\cite{Aghanim:2015xee} and \cite{Ade:2015zua}, respectively), 
 so for simplicity we do not include them in this analysis.

\section{Results}\label{sec:results}

To characterize the impact of calibration errors in LSS surveys on the ISW map reconstruction, 
and the potential to mitigate these impacts, we look at multiple combinations of input maps 
with different properties. Specifically, we consider the impact of binning in redshift, 
of adding CMB intensity data, and of including additional LSS information from another survey. For each of these studies, we examine two limiting cases for the estimator. The best case scenario,  which we will reference as $\Rtrue$, is when one perfectly models all contributions to the galaxy power, including residual calibration errors. The worst case,  referenced by $\Rfid$, is when the estimator is built out of theoretical spectra with no power from calibration errors. The power spectra in these two cases are related by  Eq.~(\ref{eq:clcalrel}).

We use the analytical $\hat\rho$ to estimate the mean reconstruction quality across a wide range of $\varc$, while performing reconstruction on simulated maps for selected values, to both verify the accuracy of $\hat\rho$ and to generate error bars for the spread of $\rho$ across simulations. 

\subsection{One survey: Binning in redshift}\label{sec:euconlyresults}

\begin{figure*}
\includegraphics[width=\linewidth]{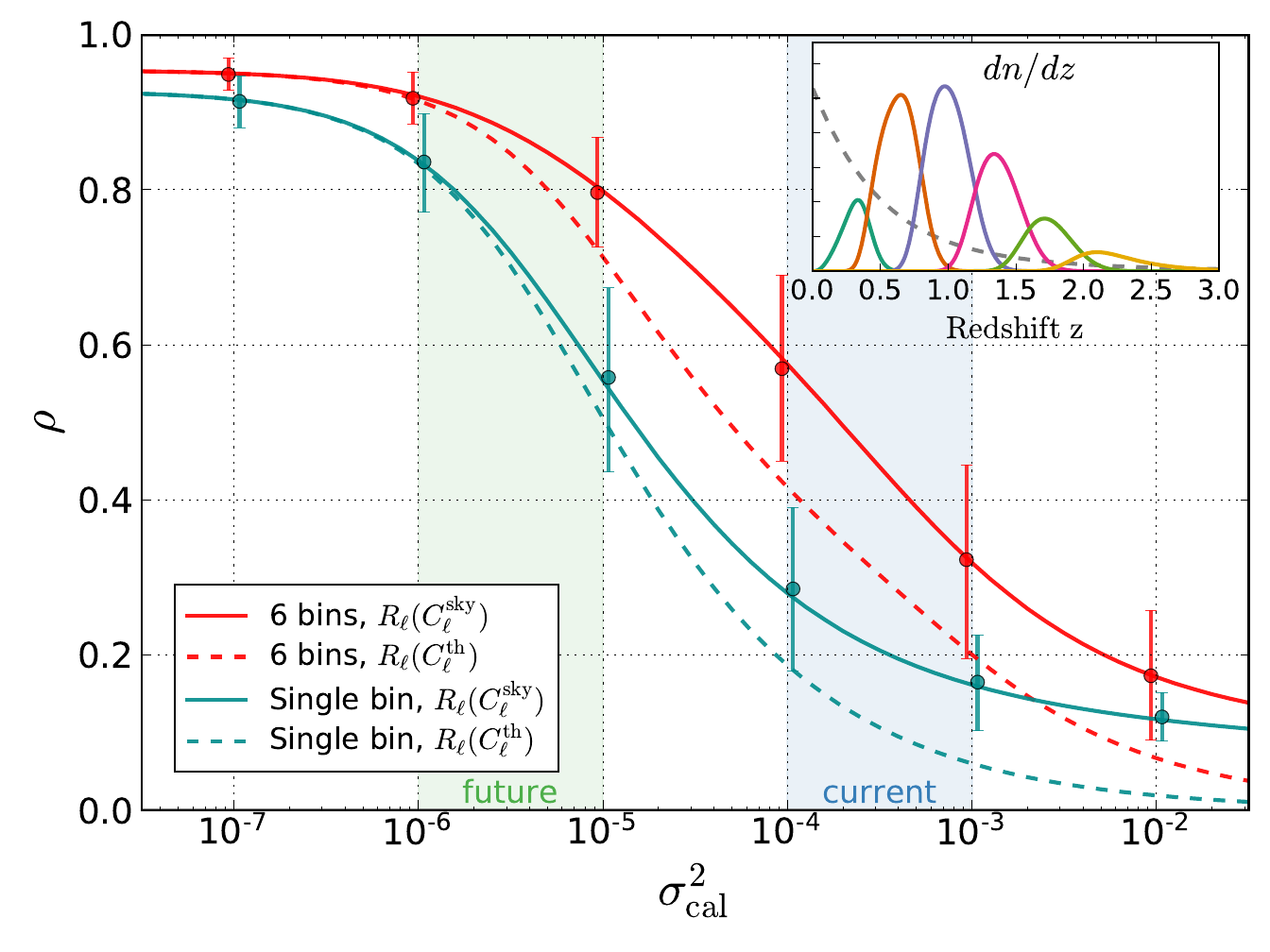}
\caption{Quality of map reconstruction $\rho$ vs.~the calibration error
  variance $\varc$ for our fiducial, Euclid-like survey. The colors of the lines indicate how tomographic information is handled, showing that splitting the survey
  into six redshift bins (red) improves the reconstruction compared to the
  single-bin case (blue).  Solid curves indicate cases when the calibration error is
  included in the ISW estimator [$\Rtrue$], while the dashed curves show the
  reconstructions in which the effects of the calibration errors are not
  included ([$\Rfid$] (see Sec.~\ref{sec:iswest} for details).  Points (offset horizontally for clarity) show the
  mean ($\bar{\rho}$) of 2,000 realizations, with error bars indicating the 68\% spread
  across realizations. The corresponding smooth curves are $\hat{\rho}$, the analytical estimate of $\bar\rho$ from Eq.~(\ref{eq:rhoest}). The inset illustrates the redshift distribution across bins overlaid with the ISW kernel in gray (reproduced from Ref.~\cite{Muir:2016veb}). The vertical, shaded regions show the approximate current and projected levels of control over residual calibration errors. Calibration errors between redshift bins are modeled as uncorrelated.}
  \label{fig:euconlyplot}
\end{figure*}

\begin{figure*}
 \includegraphics[width=\linewidth]{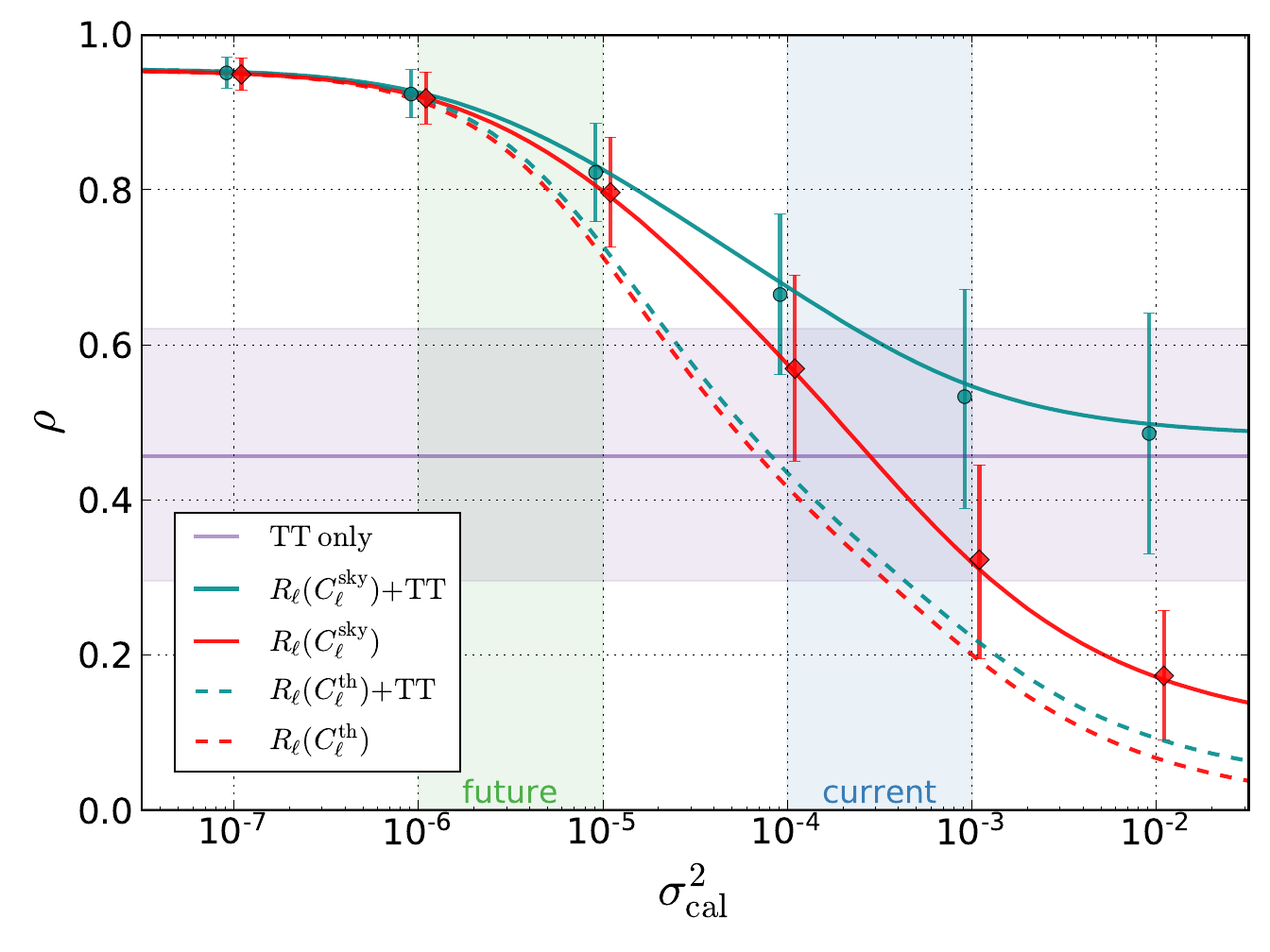}
 \caption{
  Reconstruction quality when using binned Euclid-like survey and CMB intensity data separately and in combination.
   The purple curve and shaded band show the mean and $68\%$ spread of $\rho$ 
  from simulations. As in Fig.~\ref{fig:euconlyplot}, red curves
    are results when using the binned Euclid-like survey alone, whereas blue curves are 
    the result of using both the fiducial survey and CMB intensity map. Like in 
    Fig.~\ref{fig:euconlyplot}, solid curves are for the case where calibration error power is
     correctly modeled in the estimator and dashed curves are for when they are not modeled at all.
      Neglecting the presence of calibration errors in a LSS map can actually degrade 
      the quality of the ISW reconstruction compared to using the CMB temperature alone.} 
  \label{fig:eucpluscmb}
\end{figure*}

We first consider the Euclid-like survey alone and investigate the impact of
binning in redshift on the quality of reconstructions in the presence of calibration errors.
We model calibration errors in the
binned case by adding the calibration field's power to the autopowerpower spectrum
of each bin $i$ per Eq.~(\ref{eq:clobs}): $\Cl^{i,i} \rightarrow \Cl^{i,i} + \Cl^{\rm cal}$. We do not add any power to the cross-spectra, though we test the impact of contamination in the cross-spectra in Sec.~\ref{sec:caldetails}.\footnote{In reality,
the power contribution from calibration errors will also vary somewhat across bins, depending on 
the redshift dependence of the faint-end slope of the luminosity function for the tracer population. 
We have assumed here for simplicity that the power contribution is independent of redshift.} 

 The results of this study are shown in Fig.~\ref{fig:euconlyplot}. For reference, we use a vertical shaded band to mark the level of calibration corresponding to current surveys, defined
roughly as the range bounded by the residual SDSS DR8 limiting magnitude
variations \cite{Rykoff:2015avl} and the SDSS \"uber-calibration
\cite{Padmanabhan:2007zd}. (`Future' levels are defined roughly as those
between that required to limit bias on cosmological parameters to below their
projected uncertainties and an intermediate level before bias becomes
unacceptable; see Refs.~\cite{Huterer:2012zs} and \cite{Muir:2016veb} for details.)

As shown in
Fig.~\ref{fig:euconlyplot}, splitting the survey into six redshift bins
results in significant improvement in reconstruction at all levels of
calibration error. This improvement is comparable to reducing $\varc$
of the single-bin case by a factor of $10$ at `current' levels.

Tomographic information is useful because it allows galaxy counts to 
be weighted more optimally, taking advantage of the fact that the ISW effect
becomes stronger at lower redshift as dark energy becomes more dominant 
and structure growth slows. For instance, considering the expected ISW reconstruction power
from each bin when using optimal weights (i.e. the squared contribution of each term in Eq.~\ref{eq:iswest_simple}, using $\Rtrue$), 
we find that with no calibration error, the first redshift bin contributes $87\%$ as much power 
as the second bin, with subsequent bins contributing $58\%$, $31\%$, $15\%$ and $10\%$ as much power, respectively.
There is additional benefit to binning when calibration errors are considered. Since the low-redshift bins have a higher clustering signal than the high-redshift bins,
they are less impacted by the same level of calibration error. Thus, the optimal weighting 
changes depending on the level of calibration error. When calibration errors are increased to $\varc=10^{-4}$, for example,
the first bin contributes the most power to the reconstruction,
with bins $2-6$ only contributing  $39\%$, $12\%$, $4\%$, $2\%$, and $1\%$ as much power.
As we will show later, this error-level-dependent weighting will mean adding information from a shallower survey such as SPHEREx makes reconstruction
more robust against calibration errors.

The importance of accounting for calibration errors in the estimator is apparent in the difference between the dashed and solid curves, where
doing so improves $\bar\rho$ for $\varc \gtrsim 10^{-6}$, with $\Delta \bar\rho \approx 0.1-0.2$ at current levels of calibration. This improvement is
 roughly comparable to the improvement seen from binning in redshift.

Though for clarity we do not include this case in the Figure, we additionally studied the effect of using the observed, unsmoothed galaxy-galaxy power
  in the estimator
(that is, $\tilde\Cl$, power spectra extracted from map realizations rather than computed
analytically). We find that in this case $\bar\rho$ 
  converges to the same value as the $\Rtrue$ case when calibration errors are very large, but is greatly reduced from $\bar\rho$ found using either $\Rtrue$ or $\Rfid$ when calibration errors are small ($\varc \ll 10^{-5}$).  For example, for a single input map in the limit of no calibration errors,  quality reduces from $\bar{\rho}=0.93$ to $0.83$ when we switch to using observed $\tilde\Cl$'s. If
we also use the observed (unsmoothed) \textit{cross}-correlation between the LSS
map and the CMB for the galaxy-ISW term in the estimator, reconstruction
quality is further degraded to $\bar{\rho}=0.74$ in the absence of
calibration errors. This is because primary CMB anisotropies are large compared to ISW contributions, causing the measured galaxy-CMB correlation to receive relatively large noise contributions from chance correlations between LSS maps and the primordial CMB.
      
Given the significant improvement in reconstruction that binning provides,
from here forward we adopt the configuration with six tomographic bins as our
fiducial Euclid-like survey.

\subsection{Effect of adding Planck $TT$ data}\label{sec:eucpluscmb}
We now consider adding information from the Planck-like CMB
temperature map described in Sec.~\ref{sec:modsurveys}. When 
used as the only input map,  the reconstruction
 is considerably worse than that found using the ideal Euclid-like survey (Table~\ref{table:rhotable}). 
 We include it in our study, however, because any realistic study attempting to reconstruct the ISW signal will likely include CMB temperature data. Additionally, the reconstruction quality attainable with CMB temperature data alone provides a useful baseline against which to compare the performance of estimators based on LSS maps.

 With CMB temperature data alone, we find an average reconstruction quality of $\bar{\rho} = 0.46$, in
  good agreement with Ref.~\cite{Bonavera:2016hbm}. To put this into proper context, however, it is important to note  that there is a large scatter around that mean; while the average reconstruction quality
   is indicative of performance, any single realization, such as that of our own Universe,
   can vary substantially in fidelity.
 The purple band in Fig.~\ref{fig:eucpluscmb} shows the extent of this scatter for ISW reconstruction based on just the CMB map.

When CMB temperature information is combined with that from LSS maps, it significantly improves reconstruction quality,
but only if the true galaxy power spectrum $\Cl^{\rm sky}$ (including calibration error contributions) is used in the estimator, as can been seen by the behavior of the solid
curves on the right-hand side of Fig.~\ref{fig:eucpluscmb}. The blue $\rho(\varc)$ curve
describing the CMB+LSS reconstruction tracks the maximum of the curves corresponding to reconstructions using the CMB and
LSS input maps separately, shown by the purple and red curves, respectively. This occurs because the estimator down-weights the LSS survey the
more it is affected by calibration errors, converging to the $TT$-only reconstruction quality in the limit of large calibration errors. If one
does not model calibration error power contributions, however, then any improvement from combining multiple input maps is
marginal at best and can in fact result in a \textit{worse} reconstruction
than just using the CMB data alone.  This demonstrates the importance of
ensuring that the LSS $\Cl$'s used in the ISW estimator are a good fit to the
observed spectra.

\subsection{Effect of an additional LSS survey: SPHEREx-like}\label{sec:multlss}

\begin{figure*}
\includegraphics[width=\linewidth]{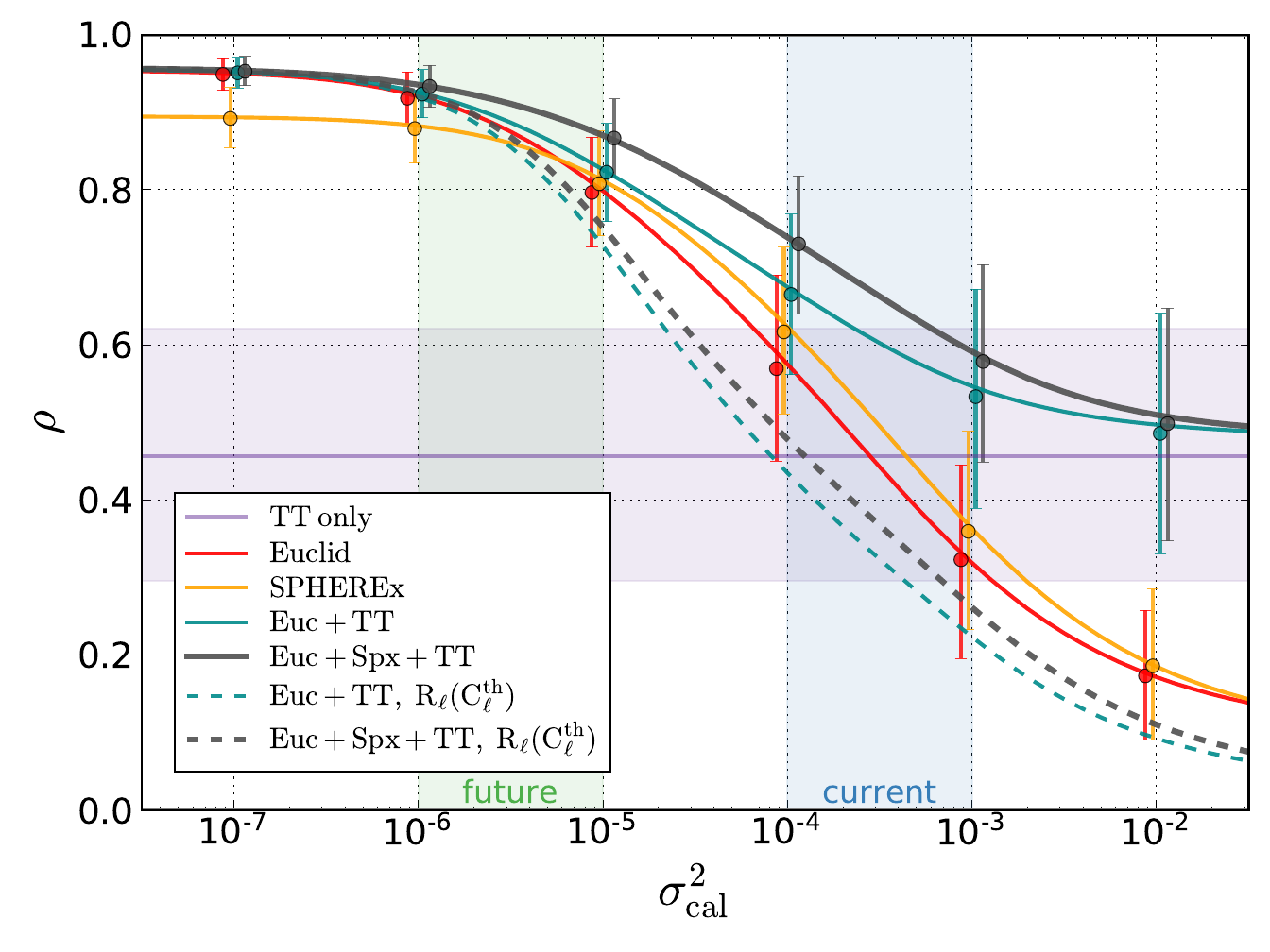}
  \caption{Comparison of ISW reconstruction quality using the LSS surveys and CMB temperature individually and in combination,
   for various levels of calibration error in the Euclid-like and 
  SPHEREx-like surveys. Colors are the same as those of Fig.~\ref{fig:eucpluscmb}. Both of the LSS surveys are split into six redshift bins 
  (see Sec.~\ref{sec:euconlyresults}), with calibration errors uncorrelated between 
  bins and surveys. The dashed curve shows the combined reconstruction if calibration errors are not included in the estimator. 
  Using LSS surveys to improve the ISW map reconstruction from 
  the CMB temperature only case requires calibration errors to be controlled
  to $\varc \lesssim 10^{-4}$. }
  \label{fig:multsurveyplot}
\end{figure*}

We now consider the addition of our fiducial six-bin SPHEREx-like survey described in Sec.~\ref{sec:spherexintro}, assuming for simplicity that it has the same level of calibration errors as the Euclid-like survey. Results are shown in Fig.~\ref{fig:multsurveyplot}.

In the limit of no calibration errors, the SPHEREx-like survey offers little additional information. In fact, adding both SPHEREx and
CMB $TT$ results in negligible improvement over the Euclid-like only case ($\Delta\bar\rho<0.003$ compared to a spread of $\sigma_{\rm Euc+Spx+TT}=0.019$).

However, by comparing the black and blue curves we see that
 including the SPHEREx-like survey does make the reconstruction somewhat more
robust against calibration errors.  The reason for this is similar to why
binning in redshift is helpful: recall that, in the case of binning, having
narrow, low-redshift bins means having some bins with higher
galaxy autopower than the unbinned case, which then 
     have less susceptibility to a given level of calibration
error.  Similarly, SPHEREx has a shallower redshift
  distribution, and thus an intrinsically higher
  clustering signal, so that it can actually provide a better reconstruction
  than the Euclid-like survey at moderate levels of calibration error.
       We would expect to see similarly increased robustness to
  calibration errors for any tracer with a larger clustering signal, including
  tracers with a larger bias.

   Finally, just as for Euclid, we find that if calibration errors are
  not accounted for in the estimator, then adding LSS data can actually result in a
  worse reconstruction than that from using CMB temperature data alone.

\renewcommand{\arraystretch}{1.5} 
\setlength{\tabcolsep}{4.5pt}     
\begin{table}
 \begin{tabular}{|c |c |c c |c c |}
 \hline
& $\Rl$ & \multicolumn{2}{c|}{$\Rfid$}& \multicolumn{2}{c|}{$\Rtrue$} \\
 \hline
 $\varc $ &  0 & $10^{-6}$ & $10^{-4}$ & $10^{-6}$ & $10^{-4}$ \\ [0.5ex] 
 \hline  \hline 
 TT & 0.46 & -  & -& -& - \\ 
 \hline
 Euclid (1 bin) & 0.92 & 0.83 & 0.19 & 0.84& 0.29\\
 \hline
 Euclid (6 bin) & 0.95 & 0.91 & 0.41& 0.92& 0.57 \\
 \hline
 SPHEREx (6 bin) & 0.89 & 0.88 & 0.52 & 0.88& 0.62\\
 \hline
 Euc + Spx + TT & 0.96 & 0.92 & 0.47 & 0.93& 0.73\\ [.5ex] 
 \hline
\end{tabular}
\caption{Mean reconstruction quality coefficients
  $\bar{\rho}$ of ISW map reconstructions for various
  combinations of input maps and select levels of calibration
  error. The second column indicates
  $\bar{\rho}$ for the case of zero calibration error.
   The following columns show the reconstruction quality for two
  nonzero values of the calibration error variance; here $\Rfid$ [$\Rtrue$] indicates
  the case where calibration errors are unaccounted [accounted] for in the
  estimator. Note, when $\varc=0$, $\Clfid=\Cltrue$. }
  \label{table:rhotable}
\end{table}

\subsection{Effect of varying calibration error properties}\label{sec:caldetails}

\begin{figure*}
\includegraphics[width=.49\linewidth]{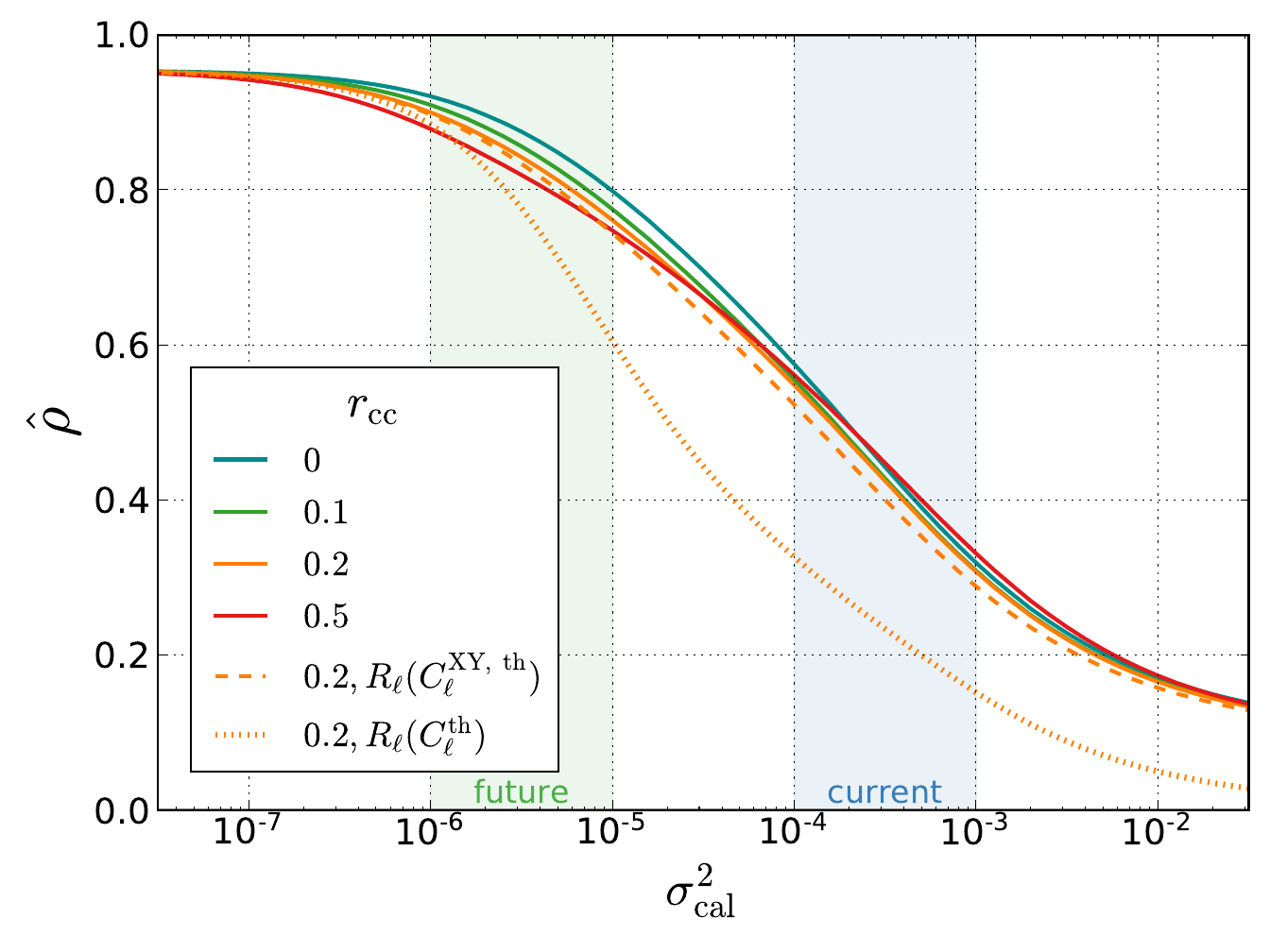}
\includegraphics[width=.49\linewidth]{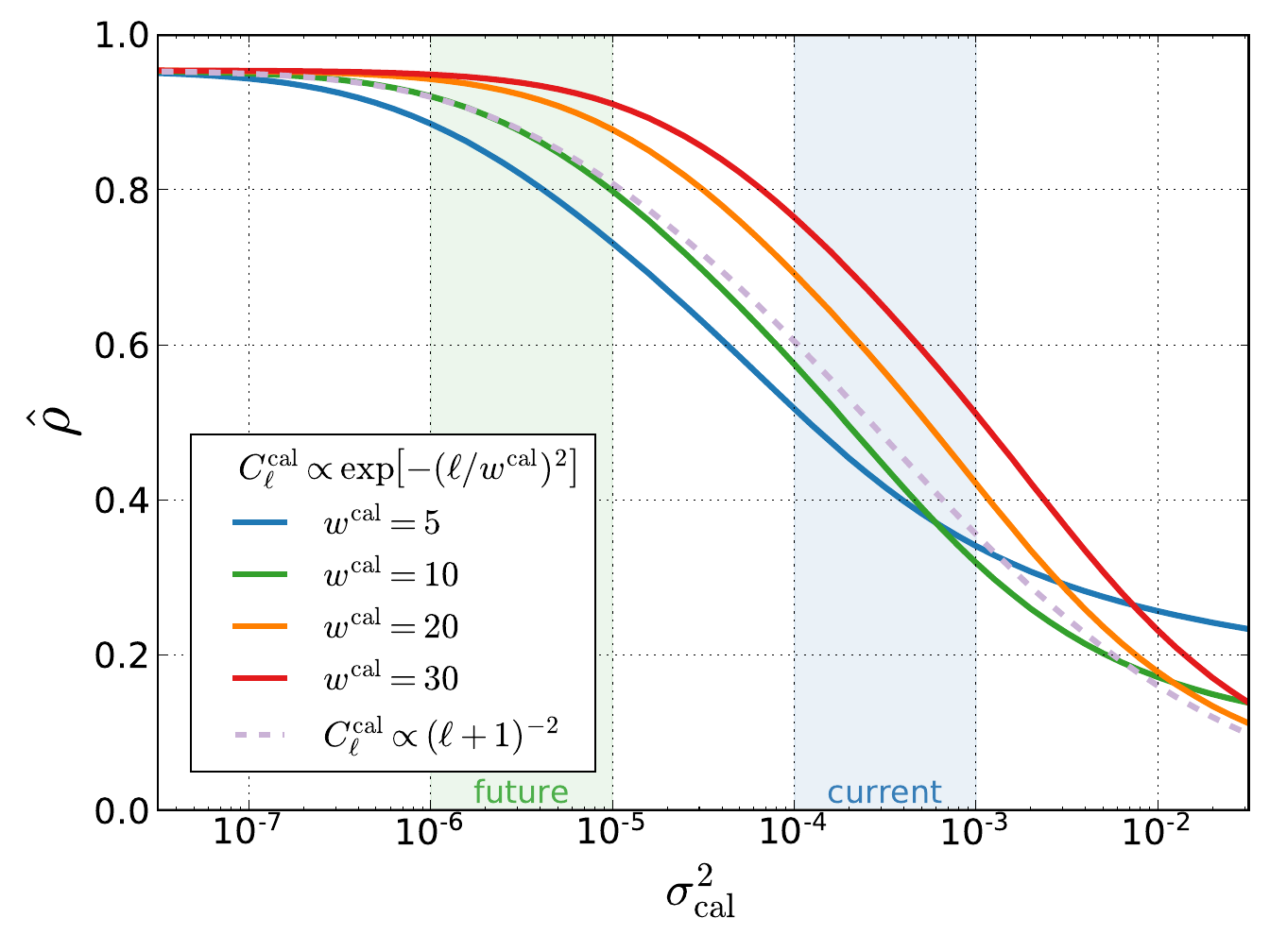}
  \caption{\textit{Left:} Effect of cross-correlation between calibration errors in different
   bins of the fiducial Euclid-like survey, given by
    $\Cl^{{\rm cal}, ij} = r_{\rm cc}\sqrt{\Cl^{{\rm cal}, ii}\Cl^{{\rm cal}, jj}}$, for bins $i\neq j$. 
    Solid curves have calibration errors accounted for in the estimator [$\Rtrue$]. The dashed curve indicates the case where only the autopower contributions of the calibration errors are accounted for in the estimator [$\Rl(\Cl^{\rm XY,th})$] and the dotted curve indicates the case where calibration errors are not accounted for at all [$\Rfid$].
    Cross-correlation of the errors results in mild degradation of the reconstruction for 
    $\varc \lesssim10^{-4}$, but otherwise has little effect as long as the auto-correlation 
    is correctly modeled in the estimator. \textit{Right:} Dependence of $\rho$ on the shape 
    of $\Cl^{cal}$. Solid curves indicate $\Cl^{cal}$ of the same form as Eq.~(\ref{eq:clcalmodel})
    but with width $w^{\rm cal}$ varied. The dashed curve indicates the case where the error spectrum
    takes the form $\Cl^{cal} \propto (\ell + 1)^{-2}$. Reconstruction fares worse when 
   calibration error power contributions are more concentrated at the largest angles, where the ISW kernel is largest. 
    In all cases, the estimator uses the true observed LSS power spectrum ($\Cltrue$).}

  \label{fig:calprops}
\end{figure*}

We now test how sensitive the results in the previous sections are to  our
assumptions about calibration errors, showing the results in Fig.~\ref{fig:calprops}.

First, the left panel shows what happens when we
vary the level of cross-correlation between the calibration errors of different
LSS maps. It is conceivable that residual calibration errors can be correlated
across different bins of a single survey, or even across different surveys,
especially if the error has an astronomical origin.  To model such
correlation, we set the level of cross-correlation between the calibration
errors of maps $i$ and $j$ using a parameter $r_{\rm cc}$, where 
\be \Cl^{{\rm cal}, ij} = r_{\rm cc}\sqrt{\Cl^{{\rm cal},
    ii}\Cl^{{\rm cal}, jj}}, \text{ for } i\neq j
\label{eq:crosscorr}
\ee
As we only consider cases where calibration errors in all maps are characterized by the same $\Cl^{\rm cal}$, this reduces to
 \be
\Cl^{{\rm cal}, ij} = r_{\rm cc}\Cl^{\rm cal},\text{ for }i\neq j. 
\label{eq:crosscorr-simp}
\ee

We consider the six-bin fiducial Euclid-like survey and find that this kind of 
correlated error results in mild degradation of the
reconstruction for $\varc \lesssim10^{-4}$, but otherwise it has little effect as
long as calibration errors are correctly modeled in the estimator [that is, $\Rtrue$ is used].

If calibration errors are \textit{not} accounted for [$\Rfid$ is used], reconstruction
suffers considerably, as shown by the dotted curve. 
We also use a dashed curve [labeled $\Rl(\Cl^{\rm XY,th})$] for the case where the
estimator filter correctly accounts for the  autopower contributions of
calibration errors but neglects the cross-power contributions. 
As seen by comparing the solid, dashed, and dotted orange curves in Fig.~\ref{fig:calprops},
reconstruction quality is far more sensitive to accurate modeling of the
calibration error contribution to the autopower than to the
cross-power.  Thus, fitting the observed autopower for each map but using theoretical cross-powers, as is done in Ref.~\cite{Barreiro:2012yh}, should 
harm the reconstruction relatively little, depending on the fitting scheme; we
find $\Delta \hat\rho \approx -0.03$ at $\varc =
10^{-4}$ for $r_{\rm cc}=0.2$, far less than the typical variation over
realizations shown in Fig.~\ref{fig:euconlyplot}.

Additionally, we study the impact of changing the shape of the calibration error power spectrum $\Clcal$, showing the results in the right panel of Fig.~\ref{fig:calprops}. We first vary the width parameter $w^{\rm cal}$ of the calibration error power spectrum $\Clcal$ given in Eq.~(\ref{eq:clcalmodel}). Results for different values of $w^{\rm cal}$ are qualitatively similar, though for fixed $\varc$, the reconstruction is less sensitive to calibration errors when $w^{\rm cal}$ is larger. The reason for this is that $\rho$ is most sensitive to contamination at the lowest multipoles, as will be discussed in Sec.~\ref{sec:SNcal}. Using a power law $\Clcal\propto(\ell+1)^{-2}$ gives results similar to our fiducial Gaussian form with $w^{\rm cal}=10$.


\begin{figure*}
\includegraphics[width=\linewidth]{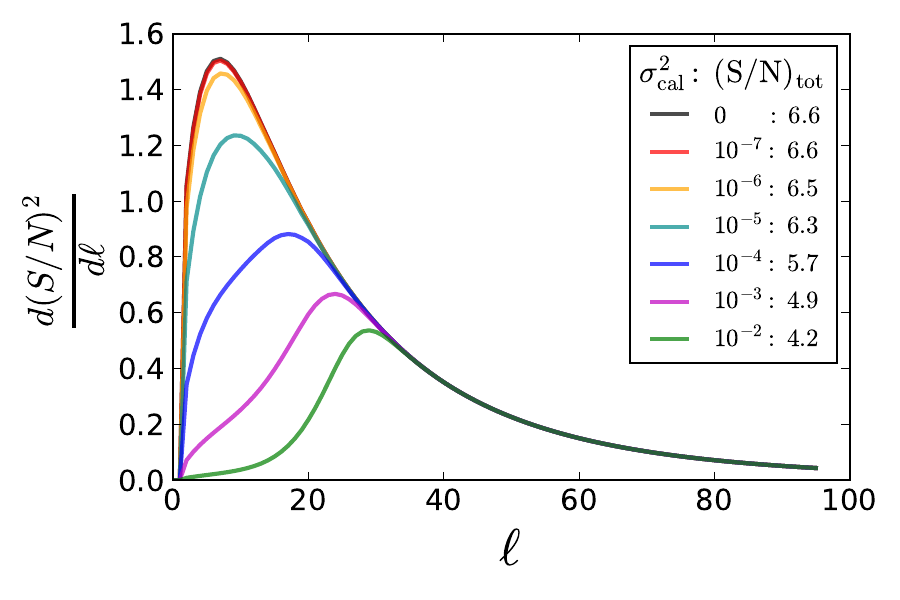}
  \caption{
  Contribution to squared signal-to-noise ratio per multipole for our fiducial 
  Euclid-like survey with varying levels of calibration error. Total combined $S/N$ for 
  each level of calibration error $\varc$ is given in the legend.}
  \label{fig:SN2l}
\end{figure*}

\section{Impact of calibration errors on S/N estimates}\label{sec:SNcal}

Given the extent to which calibration errors degrade the quality of ISW signal reconstruction, it makes sense to ask whether they also impact the signal-to-noise (S/N) of ISW detection. Detection of the ISW signal via the cross-correlation between the CMB and LSS maps has been the subject of considerable study, as it serves as an important consistency test for the presence of dark energy. The significance of detection
 varies considerably depending on the LSS tracer sample and the statistical methods used \cite{Dupe:2010zs}, as well as how systematics in the LSS data are treated~\cite{HernandezMonteagudo:2009fb,Giannantonio:2012aa,Hernandez-Monteagudo:2013vwa}.
 \citet{Hernandez-Monteagudo:2013vwa} used systematics templates to correct the
  observed power spectra for SDSS galaxies, finding a S/N loss of $\sim0.5$ if such corrections are neglected.
  \citet{Giannantonio:2012aa} introduced a low-redshift spike in the source distributions in order to reproduce the observed excess
  autopower in NVSS and SDSS catalogs and estimate that such systematics result in an uncertainty of $\Delta\textrm{S/N}\pm 0.4$. 
The most recent results come from the Planck Collaboration, which found $\sim4\sigma$ evidence for the ISW effect,
with most of the signal coming from cross-correlation of the CMB temperature
with the NVSS radio catalog and CMB lensing \cite{Ade:2015dva}.

The maximum achievable signal-to-noise can be obtained by considering
 an ideal survey that perfectly traces
the ISW (i.e. $\Cl^{gg}=\Cl^{Tg}=\Cl^{\rm ISW}$), 
resulting in a maximum S/N $\sim6-10$ for  $\Lambda$CDM cosmology \cite{Hu_Scranton,Afshordi:2004kz,Ho:2008bz,Douspis:2008xv,Giannantonio:2008zi,Dupe:2010zs,Giannantonio:2013uqa}.

Our goal is to study how calibration errors impact the significance of ISW detection. There are multiple ways one can quantify detection of the ISW effect,
including correlation detection between LSS and the CMB, template matching to an
assumed model, or model comparison. Each of these methods relies on different assumptions and
tests different statistical questions (see Ref.~\cite{Dupe:2010zs} for a detailed
review). Here we adopt the simple correlation detection statistic which quantifies the
expected deviation from a null hypothesis of no correlation between LSS ($g$) and
CMB temperature ($T$). In this formalism the S/N for ISW detection is

\be
\left(\frac{S}{N}\right)^2 \simeq \sum_\ell (\mathbf{\Cl}^{Tg})^{\ast}
  (\textbf{C}^{\rm cov}_\ell)^{-1} \mathbf{\Cl}^{Tg},
\label{eq:SNdef}
\ee
where we have assumed the multipoles contribute independently
to the S/N. Here $\mathbf{\Cl}^{Tg}$ is a vector of the ISW-LSS cross-spectra, and the covariance matrix
 elements corresponding to LSS maps $i$ and $j$ can be written as
\begin{align}
	\Cl^{{\rm cov},ij} &= \left<\Delta \Cl^{Ti} \Delta \Cl^{Tj}\right> \\
	&\simeq \frac{\Cl^{Ti}\Cl^{Tj} + \Cl^{TT} (\Cl^{ij} + \Cl^{{\rm cal},ij} + \delta_{ij} \frac{1}{\bar{n}_{ij}})} {f_{sky}(2\ell+1)}, 
\label{eq:covar}
\end{align}
where the last term in the numerator is due to shot noise and $\delta_{ij}$ is the Kronecker delta.\footnote{Strictly speaking,
  this will result in a slight underestimate of the significance, as technically the null hypothesis covariance, with
  $\Cl^{Ti}=\Cl^{Tj}=0$ in Eq.~(\ref{eq:covar}), should be used. However, as the galaxy-ISW cross-power terms are small compared to the galaxy autopower, we follow the practice in most of the literature of keeping them in the
  S/N calculation.}  Equations (\ref{eq:SNdef}) and (\ref{eq:covar}) 
  demonstrate that all cosmological tests using LSS-CMB cross-correlation are limited in their constraining power
  due to sample variance and the relatively large amplitude of the primordial CMB fluctuations. They also
  make it clear that calibration errors will reduce the significance of ISW
detection.

   We assume calibration errors
to be uncorrelated between maps, so $\Cl^{{\rm cal},ij}~\rightarrow~\delta_{ij}\Cl^{{\rm cal},ij}$.
For a single LSS map, Eq.~(\ref{eq:SNdef}) reduces to the form
\be
\left(\frac{S}{N}\right)^2 \simeq f_{sky} \sum_\ell \frac{(2 \ell+1) (\Cl^{Tg})^2} {(\Cl^{Tg})^2 + \Cl^{TT} (\Cl^{gg} + \Cl^{\rm cal} + 1/\bar{n}_g)}.
\label{eq:SNonemap}
\ee

If there are no calibration errors, we find S/N$=6.6$ for our 
Euclid-like survey, which is near the maximum\footnote{This limit can in principle be
  increased, e.g., through the method of \citet{Frommert:2008ef} in which the
  observed LSS map is used to reduce the local variance and which in our case
  brings the maximum possible S/N to $7.2$, or through the inclusion of
  polarization data as in the work by \citet{Liu:2010re}.} for this cosmology, S/N$=6.7$.
As $\varc$ increases from $0$ to current levels, the total S/N reduces to $4.9 - 5.7$, a drop of only $\sim 15\%- 30\%$. This can be seen in the S/N values listed for various $\varc$ in the legend of Fig.~\ref{fig:SN2l}. In contrast, for the same level of error, average reconstruction quality $\hat\rho$ is reduced by $40\%-60\%$. Clearly, ISW signal reconstruction is substantially more affected by calibration errors than is ISW detection significance.

The greater robustness of the total S/N to calibration errors is due to the fact that it has support at
higher multipoles.
 This is most easily illustrated in the single-map case, where
 the  contribution per multipole to the total signal-to-noise is
\be
\left(\frac{S}{N}\right)^2_\ell \equiv \frac{d\left(S/N\right)^2}{d \ell}=
(2\ell+1)\frac{(\Cl^{Tg})^2}{\Cl^{TT}\Cl^{gg} + (\Cl^{Tg})^2}.
\label{eq:SN_ell}
\ee
 Figure~\ref{fig:SN2l} shows how the contribution per multipole responds to different levels of calibration error.

As $\varc$ increases, the signal-to-noise decreases at lower
multipoles, but contributions at higher multipoles remain unchanged. These higher-multipole contributions are thus still available to contribute to the overall S/N.

Map reconstruction is more sensitive to
the largest scales. For the single-map case, this can be illustrated analytically as follows.
Using the single-map estimator from Eq.~(\ref{eq:wiener}), we can write the estimated reconstruction quality
statistic as
\begin{align}
	\hat{\rho} &=  \frac{\frac{1}{4\pi} \sum_{\ell}(2\ell+1)\left(\frac{\Cl^{Tg}}{\Cl^{gg}}\right)\Cl^{Tg}} {\sigma_{\rm ISW} \sqrt{\frac{1}{4\pi}\sum_{\ell}(2\ell+1)\left(\frac{\Cl^{Tg}}{\Cl^{gg}}\right)^2\Cl^{gg}}} \nonumber\\
	 		&=  \frac{1} {\sigma_{\rm ISW}} \sqrt{\frac{1}{4\pi}\sum_{\ell}(2\ell+1)\frac{(\Cl^{Tg})^2}{\Cl^{gg}}} \nonumber\\
			&= \frac{1} {\sigma_{\rm ISW}}\sqrt{\frac{1}{4\pi}\sum_{\ell} \left(\frac{S}{N}\right)^2_\ell \Cl^{TT}\left(1+\frac{(\Cl^{Tg})^2}{\Cl^{gg}\Cl^{TT}}\right)} \nonumber\\
			&\approx \frac{1} {\sigma_{\rm ISW}}\sqrt{\frac{1}{4\pi}\sum_{\ell}\left(\frac{S}{N}\right)^2_\ell \Cl^{TT}}.
\label{eq:rho_from_SN}
\end{align}

Here, $(S/N)^2_{\ell}$ is the quantity given by Eq.~(\ref{eq:SN_ell}) which, when summed over $\ell$, gives (S/N)$^2$. Thus, we see from Eq.\ (\ref{eq:rho_from_SN})
 that $\hat\rho$ is proportional to a total (S/N) of which the terms are weighted by $\Cl^{TT}$. Since $\Cl^{TT}$ drops sharply as $\sim\ell^{-2}$, the quality of  map reconstruction $\hat\rho$ is more impacted by large-angle calibration errors than 
the overall S/N is. This is also a primary cause for the degradation in
reconstruction quality seen when $\varc$ was concentrated
 at lower multipoles in Sec.~\ref{sec:caldetails}.

\section{Conclusions}\label{sec:conclusion}

Reconstruction of the integrated Sachs-Wolfe signal would allow, for the first time, a clean separation of the CMB temperature anisotropies into contributions from 300,000 years after the big bang and
those from some $\sim$10 billion years later. This, in turn, would allow
for a more informed assessment of the origin of the ``large-angle CMB anomalies'' and a
more complete elimination of ISW contaminants to CMB-based measurements 
of primordial non-Gaussianity. Accurate ISW reconstruction requires
wide-angle large-scale structure maps from which the gravitational potential evolution
can be inferred, but in practice, these maps are  plagued by  photometric calibration
errors due to a host of atmospheric, instrumental, and
selection-induced systematics. In our previous work, we found that the realistic
levels of residual calibration error severely degrade the accuracy of the reconstructed ISW map.

In this paper, we investigated how the effects of residual photometric calibration errors on the ISW map reconstruction 
can be mitigated by using tomographic information and by combining data from multiple surveys. 
To quantify the amount of residual calibration errors, we use their
  variance $\varc$, the square root of which is roughly equal to the
  rms \textit{magnitude} fluctuations induced by these systematics.
  
We find that for a  Euclid-like survey with a single redshift bin, to achieve
a reconstruction comparable in quality to that derived from the CMB
temperature map alone (with an average correlation between the true and reconstructed ISW maps of only $\bar{\rho} \approx 0.46$), one must limit the
variance of the calibration error field to $\varc \lesssim
10^{-5}$. 

This can be improved significantly if we exploit the 
tomographic information available by binning the LSS data in redshift
(Fig.~\ref{fig:euconlyplot}).
We also show that if the model spectra in the
estimator differ substantially from those used to generate the input maps
e.g. by using theoretical
power spectra that do not account for the observed excess autopower in the
LSS survey from calibration errors, then reconstruction quality is substantially degraded. 
 It is therefore crucial to verify that the theoretical spectra in the estimator
are a good fit to those observed or to otherwise use smoothed fits.
     
We next consider how using multiple input maps, probing different tracers and redshift
ranges, improves ISW signal reconstruction.  We find that as long as the excess power contributed by calibration errors  is adequately modeled in the estimator, the resulting
reconstruction is always better than that from either of the input
maps individually. If the excess power from calibration errors is not modeled, however, adding a map can actually degrade reconstruction
(Fig.~\ref{fig:eucpluscmb}).  The CMB temperature map 
adds information to the reconstruction at all levels of calibration error, but
is especially valuable if the LSS maps are subject to calibration errors with $\varc \gtrsim 10^{-5}$.
 Using a six-bin SPHEREx-like survey provides qualitatively similar
results to the Euclid-like one, but because it is shallower, the reconstruction is less accurate in the
limit of no calibration errors ($\rho_{\rm Spx} =0.89\pm0.04$ vs. $\rho_{\rm
  Euc}=0.95\pm0.02$, where errors indicate the $68\%$ spread across
realizations). However SPHEREx's shallower depth and therefore intrinsically
higher clustering signal becomes an asset in the presence of calibration
errors, making the survey more robust against calibration errors and leading
to a better reconstruction for $\varc \gtrsim 6\times10^{-6}$.
(Similarly,
 we would expect the increased clustering of tracers with larger bias to help mitigate the effects of 
 calibration error as well.)
Therefore, a combination of a shallower and a deeper LSS survey provides complementarity
useful for separation of calibration errors from the ISW signal and necessary for a good ISW map reconstruction in the presence of such errors. 

Using all three simulated surveys as input---Euclid, SPHEREx, and CMB temperature---significantly improves reconstruction
compared to using a single survey with current levels of
residual calibration errors, or CMB temperature data
alone. We find
that if the calibration errors can be controlled to the level of $\varc
\lesssim 10^{-6}$ ($\varc \lesssim 10^{-5}$), then the combination of Euclid,
SPHEREx, and CMB temperature maps can produce the ISW map reconstruction to an
excellent accuracy of $\rho = 0.93 \pm 0.03$ ($\rho = 0.87\pm 0.05$). This is
roughly the same level of calibration control required for future LSS surveys
to avoid biasing measurements of cosmological parameters like the
non-Gaussianity parameter $f_{NL}$ and the dark energy equation of
state.
Thus, high-quality ISW reconstructions will, in a sense, ``come for free'' with the developments needed for
cosmology constraints with next-generation surveys.
     
We additionally test the  robustness of our results against changes to the properties of the calibration error field, looking at cross-correlations between calibration
errors in different maps as well as the shape of their spectrum.  We found that
cross-correlation between the calibration errors of different tracer maps
degrades the reconstruction most for $10^{-6} \lesssim \varc \lesssim
10^{-4}$, but that this effect is relatively minor, provided the auto-correlation contribution is accounted for in the estimator
(Fig.~\ref{fig:calprops}, left).

We also find that, compared to map reconstruction accuracy, the overall significance of ISW detection is less strongly
affected by calibration errors. This is because its signal is distributed more broadly
in multipole space than that of the map reconstruction quality statistic. To
clarify this, we analytically relate $\hat{\rho}$ to the commonly used ISW detection S/N statistic  in the case of a single input LSS map and show
that $\hat{\rho}$ is weighted by an additional factor of $\Cl^{TT}$, demonstrating map reconstruction's greater sensitivity to the largest scales [Eq.~(\ref{eq:rho_from_SN})]. 

 As an extension to this work, one could study how the inclusion of CMB lensing and polarization data
can improve ISW map reconstruction,
provided the systematics present in those data sets could be
sufficiently accurately modeled.  The results of \citet{Bonavera:2016hbm}
indicate that the use of lensing as input can contribute significantly to
reconstruction quality, but they also show that current noise levels limit its
effectiveness.  Notably, the residual lensing systematics at $\ell \leq 8$
 present a challenge, 
as this is where the ISW signal is strongest, and we expect these systematics
to affect reconstruction with CMB lensing and polarization in a
manner broadly similar to unaccounted for calibration errors in LSS maps at those scales.

Further work could also be performed to concretely explore how to best
approximate the `best case' reconstruction scenario, wherein calibration
errors are fully accounted for, by using real input data. Here we have only
characterized the limiting cases where the residual calibration error contribution to the 
LSS power is fully known or fully unknown, and have not addressed intermediate cases
where they are partially accounted for.

Finally, we have only worked in the
full-sky case whereas real data will necessarily have only partial sky
coverage. Others have already shown that incomplete sky coverage only very
minorly degrades reconstruction quality for areas covered by the input
data sets \cite{Bonavera:2016hbm}, and we do not expect this to change in the
presence of calibration errors. Nevertheless, a full analysis that attempts to
make predictions for real surveys should take the actual sky coverage and
survey-specific systematics into account.

Even with these considerations,
it is clear that accounting for the presence of residual calibration errors in LSS surveys is a critical step
for any reconstruction of the ISW map, as their presence and treatment impact both the 
survey characteristics and set of input maps that produce the optimal reconstruction.  
 
\acknowledgments The authors have been supported by DOE under Contract No. DE-FG02-95ER40899. D. H. has also been supported by NASA under Contract No. 14-ATP14-0005. J. M. has been supported by the Rackham Graduate School through a Predoctoral Fellowship.


\appendix
\section{Equivalence with Other Estimators}\label{app:other_est}

We now demonstrate the equivalence between our estimator  for the ISW map coefficients
  $\hat{a}_{\ell m}^{\rm ISW}$, and the estimators proposed by \citet{Manzotti:2014kta}
  and \citet{Barreiro:2008sn}.
  
Our estimator in Eq.~(\ref{eq:iswest_simple}) is based on a version of the likelihood from \citet{Manzotti:2014kta} that has been reformulated to handle observed CMB maps like any other input map. \citet{Manzotti:2014kta}
  derive their estimator using the likelihood
\begin{align}
\label{eq:mdlike}
\mathcal{L}(T^{\rm ISW})&\propto \frac{1}{\sqrt{{\rm det}(CD)}}
\times \exp\left\{-\frac{1}{2}\vec{d}^TD^{-1}\vec{d}\right\} \\
&\times \exp\left\{-\frac{1}{2}\left(T^{\rm obs} - T^{\rm
  ISW}\right)C^{-1}\left(T^{\rm obs} - T^{\rm ISW}\right)\right\},
\nonumber
\end{align}
where $C\equiv C^p+C^n$ is the angular power spectrum of the primordial $C^p$ and noise $C^n$ contributions to CMB temperature fluctuations, $\vec{d}$ is a vector of ISW and LSS tracer maps, and $D$ is the covariance matrix between the ISW and LSS tracers (see \cite{Manzotti:2014kta} Eqs. (4-6)), with ISW maps associated with the first (1) index.
This likelihood is a product of the independent likelihoods for $(T^{\rm obs}-T^{\rm ISW})$ and for the input maps in $d$. 

Instead of explicitly including independent terms for the primordial CMB and LSS tracers (which are assumed to have no cross-correlation), we include the total observed CMB temperature,
\be
T^{\rm obs} = T^p + T^{\rm ISW},
\ee
where $T^p$ includes both the primordial CMB temperature as well as any instrumental noise terms.
We then expand the data vector to include $T^{\rm obs}$: 
\be
  \mathbf{d}_{\ell m} = (\alm^{\rm ISW},\glm^1,\dots,\glm^n) \rightarrow (\alm^{\rm ISW},\glm^1,\dots,\glm^n, \alm^{\rm obs}), \nonumber
\ee
 with $\alm$ indicating spherical components of ISW and CMB temperature fluctuations and $\glm$ indicating components of LSS overdensity.
The covariance matrix is similarly expanded to account for the cross-correlation of $T^\mathrm{obs}$ with the ISW and LSS tracers
\begin{align}
    D_{\ell} 
    &\rightarrow
    \left(\begin{array}{ccccc}
      C_{\ell}^{\rm{ISW,ISW}}&C_{\ell}^{\rm{ISW, }1} &\cdots &C_{\ell}^{\rm{ISW, }\textit{n}}&C_{\ell}^{\rm{ISW,obs}}\\
      C_{\ell}^{1\rm{,ISW}}&C_{\ell}^{1,1} &\cdots &C_{\ell}^{1,n}&C_{\ell}^{1\rm{,obs}}\\
      \vdots&\vdots &\ddots &\vdots\\
      C_{\ell}^{n\rm{,ISW}}&C_{\ell}^{n,1}&\cdots &C_{\ell}^{n,n}&C_{\ell}^{n\rm{,obs}}\\
      C_{\ell}^{\rm{obs, ISW}}&C_{\ell}^{1,\rm{obs}}&\cdots &C_{\ell}^{n,\rm{obs}}&C_{\ell}^{\rm{obs,obs}}\\
    \end{array}\right).
\end{align}

Assuming that at the scales we consider the observed CMB is cross-correlated with other LSS tracers only through the ISW, we have 
\begin{align}
C_{\ell}^{\rm{obs, ISW}} &= C_{\ell}^{\rm{ISW, ISW}} \nonumber\\
C_{\ell}^{\rm{obs, LSS}_i} &= C_{\ell}^{\rm{ISW, LSS}_i}, \\
C_{\ell}^{\rm{obs,obs}} &= C_{\ell}^{p,p} + C_{\ell}^{\rm{ISW, ISW}},\nonumber
\end{align}
assuming there is no residual cross-correlation between the primordial and late-time CMB.
Maximizing the resulting likelihood
\be
\mathcal{L}(T^{\rm ISW})\propto \frac{1}{\sqrt{{\rm det}(D)}}
\times \exp\left\{-\frac{1}{2}\vec{d}^TD^{-1}\vec{d}\right\},
\label{eq:uslike}
\ee
gives the optimal estimator given in Sec. \ref{sec:iswest}.

To show that this is equivalent to the estimator derived from Eq.~(\ref{eq:mdlike}), we focus on the case of using CMB temperature and a single LSS tracer as input maps. For compactness, and to make the connections with other ISW estimators in the literature more apparent, we adopt  notation from ~\citet{Barreiro:2008sn}, where $\ISW$, $\gal$, and $\cmb$ indicate the ISW, LSS tracer, and observed CMB temperature signals, respectively. We then have 
\be
  \mathbf{d}_{\ell m} = (\alm^\ISW,\glm^\gal, \alm^\cmb),
\ee
and
\begin{align}
   D_{\ell} &=  \left(\begin{array}{ccc}
      \Clss&\Clgs &\Clss\\
      \Clgs&\Clgg &\Clgs\\
      \Clss&\Clgs&\Cltt\\
    \end{array}\right).
\end{align}

From Eqs.~(\ref{eq:iswest_simple}) and (\ref{eq:R_def}) our estimator gives

\begin{align}
 \hat{a}_{\ell m}^{\ISW} &= \frac{-1}{[\Dl^{-1}]_{11}}\left([\Dl^{-1}]_{12}~\glm^\gal + [\Dl^{-1}]_{13}~\alm^\cmb\right)\nonumber\\
 &= \left(\frac{\Clgs(\Cltt-\Clss)}{\Clgg \Cltt - (\Clgs)^2}\right)\glm^\gal  \\
 &+ \left(\frac{\Clss\Clgg-(\Clgs)^2}{\Clgg \Cltt -
   (\Clgs)^2}\right)\alm^\cmb. \nonumber
\label{eq:2binestus}
\end{align}

We now calculate the estimator of~\citet{Manzotti:2014kta}. Denoting their covariance matrix without the CMB as $\Dl'$, we use their Eq.~(9)
\begin{align}
 \hat{a}_{\ell m}^{\ISW, \mathrm{MD}} &=
 \left([C_{\ell}^{pp}]^{-1} + [D'^{-1}]_{11}\right)^{-1}\nonumber\\
 &\times \left(-[\Dl'^{-1}]_{12}~\glm^\gal + [C_{\ell}^{pp}]^{-1}\alm^\cmb\right)\\
   &= \left(\frac{1}{\Clpp} + \frac{\Clgg}{\mathrm{det}|\Dl'|}\right)^{-1}
    \left[\frac{\Clgs}{\mathrm{det}|\Dl'|}\glm^\gal + \frac{1}{\Clpp}\alm^\cmb\right] \nonumber \\
   &= \left(\frac{1}{\mathrm{det}|\Dl'| + \Clgg\Clpp}\right)
    \left[\Clpp\Clgs\glm^\gal + (\mathrm{det}|\Dl'|)\alm^\cmb\right] \nonumber
\end{align}

Expanding the determinant,
 we find
\begin{equation}
\begin{aligned}  
\hat{a}_{\ell m}^{\ISW, \mathrm{MD}} &=
   \left(\frac{\Clgs\Clpp}{{\Clgg(\Clss+\Clpp) - (\Clgs)^2}}\right)\glm^\gal \\
   &+ \left(\frac{\Clss\Clgg-(\Clgs)^2}{\Clgg(\Clss+\Clpp) - (\Clgs)^2}\right) \alm^\cmb,
%
\end{aligned}
\end{equation}
which, using the relation $\Cltt = \Clpp + \Clss$, is equivalent to the estimator given by Eq.~(\ref{eq:2binestus}). 
\\
\\

This is also equivalent to the estimator proposed in~\citet{Barreiro:2008sn}, which uses
 the Cholesky decomposition ($L$) of the covariance matrix $\Dl''$ (denoted $\Cl$ in Ref.~\cite{Barreiro:2008sn}).
\begin{align}
 \Dl'' &=
\begin{bmatrix}
    \Clgg  & \Clgs  \\
    \Clgs & \Clss \\
\end{bmatrix}
 	= L_\ell L_\ell^T.
\end{align}
Note that here the ISW index is last instead of first, in contrast to the covariance matrices $\Dl$ and $\Dl'$ used previously.
The Cholesky decomposition is written as 
\be
L_\ell = 
\begin{bmatrix}
    \sqrt{\Clgg}  & 0  \\
    \frac{\Clgs}{\sqrt{\Clgg}} & \sqrt{\Clss-\frac{(\Clgs)^2}{\Clgg}}\\
\end{bmatrix}.
\label{eq:cholesky}
\ee

Eqs.~(8) and (9) from Ref.~\cite{Barreiro:2008sn} give the ISW estimate (their $\hat{s}_{\ell m}$) as
\begin{align*}
\hat{a}_{\ell m}^{\ISW, \mathrm{B08}} &= \frac{L_{12}}{L_{11}}\left(1-\frac{L_{22}^2}{L_{22}^2 + \Clpp}\right) \glm^\gal + \frac{L_{22}^2}{L_{22}^2 + \Clpp} \alm^\cmb\\
\end{align*}
%
where we have suppressed the $\ell$-dependence of $L$ and combined their observed ISW signal ($s_{\ell m}$) and noise ($n_{\ell m}$) terms into the single term $\alm^{\cmb}$. We use  $\Clpp$ to denote the combined power of noise and the primordial CMB, in keeping with the notation above.
Plugging this into Eq.~(\ref{eq:cholesky}), we obtain
\begin{align}
\hat{a}_{\ell m}^{\ISW, \mathrm{B08}} &= \frac{\Clgs}{\Clgg}\left(\frac{\Clpp}{(\Clss + \Clpp)-(\Clgs)^2/\Clgg}\right) \glm^\gal \nonumber \\
    &+ \frac{\Clss-(\Clgs)^2/\Clgg}{(\Clss+\Clpp) - (\Clgs)^2/\Clgg} \alm^\cmb\nonumber\\[0.2cm]    
 &= \left(\frac{\Clgs(\Cltt-\Clss)}{\Clgg \Cltt - (\Clgs)^2}\right)\glm^\gal \nonumber \\
 &+ \left(\frac{\Clss\Clgg-(\Clgs)^2}{\Clgg \Cltt - (\Clgs)^2}\right)\alm^\cmb,
\end{align}
which is the same as Eq.~(\ref{eq:2binestus}).

\section{Estimating $\rho$ with $\Rl(\tilde\Cl)$}\label{sec:appapprox}

Here we show why using raw pseudo-$\Cl$'s ($\tilde\Cl$) in the estimator results in a degraded reconstruction, for which $\bar{\rho}$ is not
well approximated by $\hat{\rho}$ (Eq.~(\ref{eq:rhoest})).

For a given realization, $\rho$ is constructed from the covariance between the true and reconstructed ISW maps (Cov$(T^{\rm ISW},T^{\rm rec})$, i.e. the numerator in Eq.~(\ref{eq:rhoraw})) normalized by the square root of the individual variances of the true and reconstructed ISW maps ($\sigma_{\rm True}^2$ and $\sigma_{\rm rec}^2$, respectively). We therefore focus on how using realization-specific $\tilde\Cl$'s in the estimator filter $\Rl$ affects the individual $\Cl$ contributions to $\sigma_{\rm rec}^2$ and Cov$(T^{\rm ISW},T^{\rm rec})$ ($\sigma_{\rm ISW}^2$ is unaffected by our choice of $\Rl$). For simplicity, we work with a single input map.

If the ISW estimator filter $\Rl$ is constructed from analytically computed model $\Cl$'s, the power spectrum of the ISW map for a given realization will be
\begin{equation}
\begin{aligned}
  \tilde\Cl^{\rm rec-rec, th} &= \Rl^2(\Clfid)\tilde\Cl^{\rm gal-gal},\\
  &= \left(\frac{\Cl^{\rm ISW-gal}}{\Cl^{\rm gal-gal}}\right)^2\,\tilde\Cl^{\rm gal-gal}.
\end{aligned}
\end{equation}
We add the superscript ``th'' to distinguish this reconstructed ISW power spectrum from the one where the filter $\Rl$ is built from $\tilde\Cl$'s, which will be discussed shortly. The expectation value for this over many realizations is
\begin{equation}
\begin{aligned}  
  \left\langle \tilde\Cl^{\rm rec-rec, th}\right\rangle 
  &= \left(\frac{\Cl^{\rm ISW-gal}}{\Cl^{\rm gal-gal}}\right)^2\,\left\langle\tilde\Cl^{\rm gal-gal}\right\rangle,\\
  &=\frac{\left(\Cl^{\rm ISW-gal}\right)^2}{\Cl^{\rm gal-gal}}.
\end{aligned}
\end{equation}
Now let us look at the behavior of the reconstructed ISW power when the galaxy autopower spectra in the estimator filter are extracted from the observed maps. Denoting this version of the filter by
\be
\tilde\Rl\equiv \frac{\Cl^{\rm ISW-gal}}{\tilde\Cl^{\rm gal-gal}},
 \ee
we write 

\begin{equation}
\begin{aligned}  
  \tClrec &= \tilde\Rl^2\tilde\Cl^{\rm gal-gal},\\
  &=\left(\Cl^{\rm ISW-gal}\right)^2\left(\frac{1}{\tilde\Cl^{\rm gal-gal}}\right). 
\label{eq:Rlsingle}
\end{aligned}
\end{equation}

Because the  measured $\tilde\Cl^{\rm gal-gal}$ appears in the denominator of this expression, taking its expectation value over many realizations is somewhat more complicated. To do so we use the fact that $(2\ell+1)\tilde\Cl^{\rm gal-gal}$ is $\chi^2$-distributed with $2\ell+1$ degrees of freedom. This means 
 $\tClrec/(2\ell+1)$ follows an inverse-$\chi^2$ distribution, with an expectation value\footnote{We refer the reader to Refs.~\cite{Hartlap:2006kj} and \cite{Taylor:2012kz} for discussions of the bias introduced when inverting an estimator, with implications specifically for estimating the inverse covariance matrix.} 
\be
\left< \frac{\tClrec}{2\ell+1} \right> = \frac{1}{2\ell -1}\frac{\left(\Cl^{\rm ISW-gal}\right)^2}{\Cl^{\rm gal-gal}}.
\ee
Therefore the average reconstructed power is
\begin{equation}
\begin{aligned}  
\left<\tilde\Cl^{\rm rec-rec}\right> &= \frac{2\ell+1}{2\ell-1}\left<\tilde\Cl^{\rm rec-rec, th}\right> \label{eq:recrecanal} \\
 &= \left<\tilde\Cl^{\rm rec-rec, th}\right>\left(1 + \frac{2}{2\ell-1}\right).
\end{aligned}
\end{equation}
Because $\langle\tClrec\rangle$ is strictly positive, this increased power results in an increase in the total variance of the reconstruction map $\tilde \sigma_{\rm rec}^2$ compared to that from the theory-only filter reconstruction $\sigma_{\rm rec, th}^2$

\begin{equation}
\begin{aligned}  
  \langle\tilde \sigma_{\rm rec}^2\rangle &=\frac{1}{4\pi} \sum_{\ell}\,(2\ell+1)\,\left<\tClrec\right> \\
  &= \sigma_{\rm rec, th}^2 + \frac{1}{4\pi} \sum_{\ell}\,(2\ell+1)\,\frac{\left<\tilde\Cl^{\rm rec-rec, th}\right>}{\ell-1/2},
\end{aligned}
\end{equation}

In contrast, we find the average cross-power $\langle\tilde\Cl^{\rm ISW-rec}\rangle$ 
 between reconstructed and true ISW maps remains unchanged. The increased power of the reconstruction thus results in a net decrease in $\left<\rho\right>$, per Eq.~(\ref{eq:rhoraw}), and hence is not well approximated by simply substituting the theory $\Cl$, as is done to compute $\hat \rho$.
 Additionally, this suggests that a simple scaling of $\Rl$ in order to ``debias" the reconstruction will not improve $\rho$.

To understand why the cross-power does not increase, we again use the observed galaxy autopower in the estimator and approximate its expectation value. The cross-power is given by
\begin{equation}
\begin{aligned}  
\tilde\Cl^{\rm ISW-rec} &= \tilde\Rl \tilde\Cl^{\rm ISW-gal}\\
  &= \left(\frac{\Cl^{\rm ISW-gal}}{\tilde\Cl^{\rm gal-gal}}\right) \tilde\Cl^{\rm ISW-gal}.
  \label{eq:singleClrecisw}
\end{aligned}
\end{equation}
Here we have a quotient of two \textit{non}-independent $\chi^2$ random variables. 
Generically, we can approximate the average of a function of two random variables $X$ and $Y$ through a second-order Taylor expansion about the mean of each $(\mu_X, \mu_Y)$:

\begin{align*}
\left< f(X,Y)\right> &\approx f(\mu_X,\mu_Y) + \frac{1}{2}f''_{XX}(\mu_X,\mu_Y)\left<(X-\mu_X)^2\right> \\
&+ f''_{XY}(\mu_X,\mu_Y)\left<(X-\mu_X)(Y-\mu_Y)\right> \\
&+ \frac{1}{2}f''_{YY}(\mu_X,\mu_Y)\left<(Y-\mu_Y)^2\right>,
\end{align*}
where a prime indicates a derivative with respect to the respective subscripted variable. 
By taking $f(X,Y)$ to be $\tilde\Cl^{\rm ISW-rec}$, of the form $X/Y$, then from Eq.~(\ref{eq:singleClrecisw}) we can approximate the mean cross-power to be 

\begin{align*}
& \left<\tilde\Cl^{\rm ISW-rec}\right> \approx \Cl^{\rm ISW-gal}\frac{\left<\tilde\Cl^{\rm ISW-gal}\right>}{\left<\tilde\Cl^{\rm gal-gal}\right>}\\
&\times \left(1 - \frac{\text{Cov}(\tilde\Cl^{\rm ISW-gal}, \tilde\Cl^{\rm gal-gal})}{\left<\tilde\Cl^{\rm ISW-gal}\right>\left<\tilde\Cl^{\rm gal-gal}\right>} + \frac{\text{Var}(\tilde\Cl^{\rm gal-gal})}{\left<\tilde\Cl^{\rm gal-gal}\right>^2}\right)\nonumber\\[0.2cm]
&= \Cl^{\rm ISW-gal}\frac{\left<\tilde\Cl^{\rm ISW-gal}\right>}{\left<\tilde\Cl^{\rm gal-gal}\right>}\left(1 - \frac{2}{2\ell+1} + \frac{2}{2\ell+1}\right),
\end{align*}
where we used
\begin{align*}
\mathrm{Var}\left(\tilde\Cl^{\rm gal-gal}\right) &= \frac{2}{2\ell+1}\Cl^{\rm gal-gal}, \\
\mathrm{Cov}\left(\tilde\Cl^{\rm gal-gal},\tilde\Cl^{\rm ISW-gal}\right) &= \frac{2}{2\ell+1}\Cl^{\rm gal-gal}\Cl^{\rm ISW-gal}.
\end{align*}
The corrective terms vanish and we find  

\begin{equation}
\begin{aligned}
\left<\tilde\Cl^{\rm ISW-rec}\right> &\approx \Cl^{\rm ISW-gal}\frac{\left<\tilde\Cl^{\rm ISW-gal}\right>}{\left<\tilde\Cl^{\rm gal-gal}\right>} \\
&=\left<\tilde\Cl^{\mathrm{ISW-rec, th}}\right>.
\end{aligned}
\end{equation}

Then on average, the cross-power between the true and reconstructed ISW maps is unchanged from the theory case.
Since the multipoles are independent, this means the total covariance between the true and reconstructed ISW maps is unchanged as well: 
\begin{align*}
\left<\tilde{\mathrm{Cov}}\left(T^{\rm ISW},T^{\rm rec}\right)\right> &=\frac{1}{4\pi} \sum_{\ell}\,(2\ell+1)\,\left<\tilde\Cl^{\mathrm{ISW-rec}}\right> \\
		&=\frac{1}{4\pi} \sum_{\ell}\,(2\ell+1)\,\left<\tilde\Cl^{\mathrm{ISW-rec, th}}\right> 
\end{align*}

While for the autopower $\tilde\Cl^{\rm rec-rec}$ we were able to derive an analytical result, a similar 
Taylor expansion treatment to the same order as the cross-power results in an additive correction of $2/(2\ell+1)$, or
\be
\left<\tilde\Cl^{\rm rec-rec}\right>\approx \left(\frac{2\ell+3}{2\ell+1}\right)\left<\Cl^{\mathrm{rec-rec, th}}\right>,
\ee
which is a good approximation to the analytical result found in Eq.~(\ref{eq:recrecanal}).

\bibliography{weaverdyckmuirhuterer_iswrec_paper_2017_prd_v3}{}
\end{document}